# Max-Min SINR in Large-Scale Single-Cell MU-MIMO: Asymptotic Analysis and Low Complexity Transceivers

Houssem Sifaou, *Student Member, IEEE*, Abla Kammoun, *Member, IEEE*, Luca Sanguinetti, *Senior Member, IEEE*, Mérouane Debbah, *Fellow, IEEE* and Mohamed-Slim Alouini, *Fellow, IEEE*

*Abstract*—This work focuses on the downlink and uplink of large-scale single-cell MU-MIMO systems in which the base station (BS) endowed with $M$ antennas communicates with $K$ single-antenna user equipments (UEs). Particularly, we aim at reducing the complexity of the linear precoder and receiver that maximize the minimum signal-to-interference-plus-noise ratio subject to a given power constraint. To this end, we consider the asymptotic regime in which $M$ and $K$ grow large with a given ratio. Tools from random matrix theory (RMT) are then used to compute, in closed form, accurate approximations for the parameters of the optimal precoder and receiver, when imperfect channel state information (modeled by the generic Gauss-Markov formulation form) is available at the BS. The asymptotic analysis allows us to derive the asymptotically optimal linear precoder and receiver that are characterized by a lower complexity (due to the dependence on the large scale components of the channel) and, possibly, by a better resilience to imperfect channel state information. However, the implementation of both is still challenging as it requires fast inversions of large matrices in every coherence period. To overcome this issue, we apply the truncated polynomial expansion (TPE) technique to the precoding and receiving vector of each UE and make use of RMT to determine the optimal weighting coefficients on a per-UE basis that asymptotically solve the max-min SINR problem. Numerical results are used to validate the asymptotic analysis in the finite system regime and to show that the proposed TPE transceivers efficiently mimic the optimal ones, while requiring much lower computational complexity.

## I. INTRODUCTION

Large-scale multiple-input multiple-output (MIMO) systems, also known as massive MIMO systems, are considered as a promising technique for next generations of wireless communication networks [1]–[4]. The massive MIMO technology aims at evolving the conventional base stations (BSs) by using arrays with a hundred or more small dipole antennas. This allows for coherent multi-user MIMO transmission where tens of users can be multiplexed in both uplink (UL) and downlink (DL) of each cell. It is worth observing that, contrary to what the name "massive" suggests, massive MIMO arrays are rather compact; 160 dual-polarized antennas at 3.7 GHz fit into the form factor of a flat-screen television [5].

The problem of designing precoding and receiving techniques for massive MIMO systems is receiving a lot of attention. Among the different optimization criteria, we distinguish the transmit power minimization [6]–[8] and the maximization of the minimum SINR [9], [10]. The latter is the focus of this work. In particular, we consider the case of a single-cell large-scale multi-user (MU) MIMO system in which the BS makes use of $M$ antennas in order to communicate with $K$ single-antenna user equipments (UEs). Under the assumption of perfect channel state information (CSI) at the BS, it is shown in [9] that the optimal linear precoder (OLP) for the max-min SINR problem is closely related to the optimal linear receiver (OLR), as it can be computed by exploiting the UL-DL duality principle. The latter allows to convert the DL optimization problem into its equivalent counterpart in the dual UL variables. The OLP is then found in the form of a fixed-point problem whose solution corresponds to the powers allocated to the UEs in the dual UL network. Although computationally feasible, the above approach does not provide any insight into the structure of both OLP and OLR.

To solve the above issue, we follow the same approach as in recent works [11]–[14] (among many others). Particularly, we consider the asymptotic regime in which $M$ and $K$ grow large with bounded ratio, which allows us to leverage recent results from random matrix theory. The analysis is performed under the assumption of imperfect CSI at the BS, which is modeled by the generic Gauss-Markov formulation form (see for example [15]). Under imperfect CSI, the OLP and OLR derived in [9] are no longer optimal [16]. This is clearly unveiled by the large system analysis, which additionally shows that the directions of the precoding and receiving vectors as well as their associated powers converge asymptotically to deterministic values depending only on the long-term channel attenuations of the UEs. In order to account for the channel estimation errors and to avoid the need for solving fixed point equations at the pace of fast fading channels, we propose the asymptotically OLP and OLR (called A-OLP and A-OLR, respectively) for which the same asymptotic directions as OLP and OLR are used but the transmit powers are computed in order to maximize the asymptotic minimum SINR under a total power constraint. We prove that A-OLP provides asymptotically better performance than OLP while OLR and A-OLR exhibit the same performance in the asymptotic regime.

Despite being reduced compared to OLP and OLR, the im-

L. Sanguinetti is with the University of Pisa, Dipartimento di Ingegneria dell'Informazione, Italy (luca.sanguinetti@unipi.it) and also with the Large Systems and Networks Group (LANEAS), CentraleSupélec, Université Paris-Saclay, 3 rue Joliot-Curie, 91192 Gif-sur-Yvette, France

M. Debbah is with the Large Systems and Networks Group (LANEAS), CentraleSupélec, Université Paris-Saclay, 3 rue Joliot-Curie, 91192 Gif-sur-Yvette, France (m.debbah@centralesupelec.fr) and also with the Mathematical and Algorithmic Sciences Lab, Huawei Technologies Co. Ltd., France (merouane.debbah@huawei.com).

H. Sifaou, A. Kammoun and M.S. Alouini are with the Computer, Electrical, and Mathematical Sciences and Engineering (CEMSE) Division, KAUST, Thuwal, Makkah Province, Saudi Arabia (e-mail: houssem.sifaou@kaust.edu.sa, abla.kammoun@kaust.edu.sa, slim.alouini@kaust.edu.sa).

The research of L. Sanguinetti and M. Debbah has been supported by the ERC Starting Grant 305123 MORE.

The work of H. Sifaou, A. Kammoun, and M. -S. Alouini was supported by the Qatar National Research Fund (a member of Qatar Foundation) under NPRP Grant NPRP 6-001-2-001 . The statements made herein are solely the responsibility of the authors.

A preliminary version of this paper has been accepted in the IEEE International Workshop on Signal Processing Advances in Wireless Communications, Edinburgh, UK, 2016.

plementation of A-OLP and A-OLR might be of prohibitively high complexity in large scale MIMO systems due to the need for computing the inverse of large matrices, whose dimensions grow with $M$ and $K$. To tackle this problem, we resort to the truncated polynomial expansion (TPE) technique, which has recently been applied to reduce the complexity of the RZF precoder in [17], [18] and the MMSE receiver in [19]–[22]. In all these aforementioned works, the TPE concept is applied using the same weighting coefficients for all UEs. This limits the number of degrees of freedom with an ensuing degradation of the maximum achievable performance. In light of this observation, we employ the TPE technique on a per-UE basis. More specifically, the TPE concept is applied to each vector of the precoding and receiving matrices rather than to the whole matrices themselves. This leads to the so-called user specific TPE (US-TPE) precoder and receiver for which approximations of the resulting SINRs are computed through asymptotic analysis. These results are then used to optimize the US-TPE parameters in order to maximize the minimum SINR over all UEs in the DL and UL. Interestingly, the optimization problem can be cast in both cases as the max-min SINR problems previously studied in [7], [9], [10]. The solution of such problem leads to a novel US-TPE precoder and receiver, which are shown by simulations to achieve almost the same performance as A-OLP and A-OLR, while requiring much lower computational complexity.

The remainder of this work is organized as follows. Next section introduces the system model and formulates the max-min SINR problem for both DL and UL. Section III deals with the large system analysis of OLP and OLR as well as with the design of both under the assumption of imperfect CSI. The proposed TPE precoder and receiver are presented in section IV. Numerical results are shown in Section V while some conclusions are drawn in Section VI.

*Notations* – Boldface lower case is used for denoting column vectors, $\mathbf{x}$, and upper case for matrices, $\mathbf{X}$, $\mathbf{X}^T$, $\mathbf{X}^H$ denote the transpose and conjugate of $\mathbf{X}$, respectively. The trace of a matrix $\mathbf{X}$ is denoted by $\operatorname{tr}(\mathbf{X})$. A circularly symmetric complex Gaussian random vector $\mathbf{x}$ is denoted $\mathbf{x} \sim \mathcal{CN}(\overline{\mathbf{x}}, \mathbf{Q})$ where $\overline{\mathbf{x}}$ is the mean and $\mathbf{Q}$ is the covariance matrix. Moreover, $\mathbf{I}_M$ denotes the $M \times M$ identity matrix and $0_{M\times 1}$ stands for the $M \times 1$ vector with all entries equal to zero. The expectation operator is denoted $\mathbb{E}[.]$. For an infinitely differentiable function $f(t)$, the $n$-th derivative at $t = t_0$ is denoted $f^{(n)}(t_0)$ and it is simply denoted by $f^{(n)}$ when $t = 0$. The operator $\operatorname{diag}\left(\{v_k\}_{k=1}^K\right)$ is the diagonal matrix having $v_1, \cdots, v_K$ as diagonal elements.

## II. SYSTEM MODEL AND PROBLEM FORMULATION

We consider the DL and UL of a single-cell multi-user MIMO system in which the BS is equipped with $M$ antennas and communicates with $K < M$ single antenna UEs. We denote by $\mathbf{h}_k \in \mathbb{C}^M$ the channel vector of UE $k$ and assume that $\mathbf{h}_k = \sqrt{\beta_k}\mathbf{z}_k$ where $\mathbf{z}_k \sim \mathcal{CN}(\mathbf{0}, \mathbf{I}_M)$ is the small-scale fading channel and $\beta_k$ accounts for the corresponding large-scale channel fading or path loss. Within the above setting, we are interested in computing the optimal linear precoder (receiver) that maximizes the minimum SINR in the DL (UL) while satisfying a total average power constraint $P_{\max}$ [1] Under the assumption of perfect CSI at the BS, the solution of this problem is well known and can be computed using different approaches based on standard convex optimization techniques. Next, for completeness we consider the DL and UL and review the optimal linear precoder and receiver structure. This will be instrumental for the asymptotic analysis performed subsequently.

### A. Downlink

Denoting by $\mathbf{g}_k \in \mathbb{C}^M$ the precoding vector associated with UE $k$, the signal received at UE $k$ can be written as

$$y_k = \mathbf{h}_k^H \mathbf{g}_k s_k + \sum_{i=1, i \neq k}^K \mathbf{h}_k^H \mathbf{g}_i s_i + n_k \qquad (1)$$

where $s_i \sim \mathcal{CN}(0,1)$ is the signal intended to UE $k$, assumed independent across $k$, $n_k \sim \mathcal{CN}(0, 1/\rho)$ accounts for the additive Gaussian noise with $\rho$ being the effective signal-to-noise ratio (SNR). The DL SINR at the $k$-th UE is:

$$\operatorname{SINR}_k^{\mathrm{dl}} = \frac{|\mathbf{h}_k^H \mathbf{g}_k|^2}{\sum_{i=1, i\neq k}^K |\mathbf{h}_k^H \mathbf{g}_i|^2 + 1/\rho} \qquad (2)$$

and the total average transmit power per UE is $\frac{1}{K}\operatorname{tr}(\mathbf{G}\mathbf{G}^H)$ where $\mathbf{G} = [\mathbf{g}_1, \cdots, \mathbf{g}_K] \in \mathbb{C}^{N \times K}$. The latter is chosen as the solution of the following max-min SINR problem:

$$\mathcal{P}_{\mathrm{dl}} : \begin{cases} \max_{\mathbf{G}} & \min_k \frac{\operatorname{SINR}_k^{\mathrm{dl}}}{\gamma_k} \\ \text{s.t.} & \frac{1}{K}\operatorname{tr}(\mathbf{G}\mathbf{G}^H) \leq P_{\max} \end{cases} \qquad (3)$$

where $\gamma_k$ is a factor reflecting the priority of UE $k$ and $P_{\max}$ is the power constraint at the BS. In [9], [16], it is shown that the column vectors of the optimal linear precoder (OLP) $\mathbf{G}^\star$ solving $\mathcal{P}_{\mathrm{dl}}$ take the form $\mathbf{g}_k^\star = \sqrt{\frac{p_k^\star}{K}} \frac{\mathbf{v}_k^\star}{\|\mathbf{v}_k^\star\|}$ with

$$\mathbf{v}_k^\star = \left( \sum_{\ell=1,\ell\neq k}^K \frac{q_\ell^\star}{K} \mathbf{h}_\ell \mathbf{h}_\ell^H + \frac{1}{\rho}\mathbf{I}_M \right)^{-1} \mathbf{h}_k \qquad (4)$$

where the scalars $\{q_k^\star\}$ are obtained as the unique positive solution to the following fixed-point equations:

$$q_k^\star = \frac{\gamma_k \tau^\star}{\frac{1}{K}\mathbf{h}_k^H \left( \sum_{k=1,\ell\neq k}^K \frac{q_\ell^\star}{K}\mathbf{h}_\ell \mathbf{h}_\ell^H + \frac{1}{\rho}\mathbf{I}_M \right)^{-1} \mathbf{h}_k} \qquad (5)$$

with $\tau^\star$ being the minimum weighted SINR given by [9]:

$$\tau^\star = \frac{KP_{\max}}{\sum_{n=1}^K \gamma_n \left( \frac{1}{K}\mathbf{h}_n^H \left( \sum_{k=1,k\neq n}^K \frac{q_k^\star}{K}\mathbf{h}_k \mathbf{h}_k^H + \frac{1}{\rho}\mathbf{I}_M \right)^{-1} \mathbf{h}_n \right)^{-1}}. \qquad (6)$$

---

[1]Throughout the paper, $P_{\max}$ refers to the power allocated to data transmission.

The optimal power coefficients $\{p_k^\star\}$ are such that the following equalities are satisfied [9]:

$$\frac{\text{SINR}_1^{\text{dl}\star}}{\gamma_1} = \cdots = \frac{\text{SINR}_K^{\text{dl}\star}}{\gamma_K} = \tau^\star \quad (7)$$

with

$$\text{SINR}_k^{\text{dl}\star} = \frac{\frac{p_k^\star}{K}\frac{|\mathbf{h}_k^H \mathbf{v}_k^\star|^2}{\|\mathbf{v}_k^\star\|^2}}{\sum_{i=1,i\neq k}^{K}\frac{p_i^\star}{K}\frac{|\mathbf{h}_k^H \mathbf{v}_i^\star|^2}{\|\mathbf{v}_i^\star\|^2} + 1/\rho}. \quad (8)$$

From the above condition, it turns out that $\mathbf{p}^\star = [p_1^\star, \cdots, p_K^\star]^T$ can be obtained as [9]:

$$\mathbf{p}^\star = \frac{\tau^\star}{\rho}\left(\mathbf{I}_K - \tau^\star \mathbf{\Gamma}\mathbf{F}\right)^{-1}\mathbf{\Gamma}\mathbf{1}_K \quad (9)$$

where $\mathbf{\Gamma} = \text{diag}\{\frac{K\gamma_1\|\mathbf{v}_1^\star\|^2}{|\mathbf{h}_1^H\mathbf{v}_1^\star|^2},\cdots,\frac{K\gamma_K\|\mathbf{v}_K^\star\|^2}{|\mathbf{h}_K^H\mathbf{v}_K^\star|^2}\}$ and $\mathbf{F} \in \mathbb{C}^{K\times K}$ has elements given by:

$$[\mathbf{F}]_{k,i} = \begin{cases} 0 & \text{if } k = i \\ \frac{1}{K}\frac{|\mathbf{h}_k^H\mathbf{v}_i^\star|^2}{\|\mathbf{v}_i^\star\|^2} & \text{if } k \neq i. \end{cases} \quad (10)$$

*B. Uplink*

From the UL-DL duality shown in [9], it follows that the vectors $\{\mathbf{v}_k^\star\}$ and $\mathbf{q}^\star = [q_1^\star,\cdots,q_K^\star]^T$ can be obtained as the solution of the following uplink max-min SINR problem:

$$\mathcal{P}_{\text{ul}}: \begin{cases} \max_{\{\mathbf{v}_k\},\mathbf{q}} & \min_k \frac{\text{SINR}_k^{\text{ul}}}{\gamma_k} \\ \text{s.t.} & \frac{1}{K}\mathbf{1}_K^T\mathbf{q} \leq P_{\max} \end{cases} \quad (11)$$

with

$$\text{SINR}_k^{\text{ul}} = \frac{\frac{q_k}{K}|\mathbf{h}_k^H\mathbf{v}_k|^2}{\mathbf{v}_k^H\left(\sum_{i=1,i\neq k}^{K}\frac{q_i}{K}\mathbf{h}_i\mathbf{h}_i^H + \frac{1}{\rho}\mathbf{I}_M\right)\mathbf{v}_k}. \quad (12)$$

From (12), it easily follows that the vector $\mathbf{v}_k$ solving $\mathcal{P}_{\text{ul}}$ coincides with the minimum-mean-square-error (MMSE) receiver [23]. Next, we refer to the solution of $\mathcal{P}_{\text{ul}}$ as the optimal linear receiver (OLR).

## III. LARGE SYSTEM ANALYSIS

As shown above, the OLP and OLR are parametrized by the scalars $\{q_k^\star\}$ and $\{p_k^\star\}$ where $\{q_k^\star\}$ need to be evaluated by an iterative procedure due to the fixed-point equations in (5) and (6). This is a computationally demanding task when $M$ and $K$ are large since the matrix inversion operation in (5) and (6) must be recomputed at every iteration. Moreover, computing $\{q_k^\star\}$ as the fixed point of (5) and (6) does not provide any insight into the optimal structure of $\{q_k^\star\}$ and consequently of $\{p_k^\star\}$ in (9). In addition, both depend directly on the channel vectors $\{\mathbf{h}_k\}$ and change at the same pace as the small-scale fading (i.e., at the order of milliseconds). To overcome these issues, we exploit the statistical distribution of $\{\mathbf{h}_k\}$ and the large values of $M, K$ (as envisioned in future networks) to compute deterministic approximations (also known as deterministic equivalents) of $\{q_k^\star\}$ and $\{p_k^\star\}$. For technical purposes, we shall consider the following assumptions:

**Assumption 1.** *We assume that both $M$ and $K$ grow large, their ratio being bounded below and above as follows:* $1 < \liminf \frac{M}{K} \leq \limsup \frac{M}{K} < \infty$.

**Assumption 2.** *The channel attenuation coefficients $\{\beta_k\}$ satisfy:* $0 < \liminf\{\beta_k\} \leq \limsup\{\beta_k\} < \infty$.

**Assumption 3.** *The power coefficients $\{p_k\}$ satisfy:* $0 < \liminf \min_i p_i < \limsup \max_i p_i < \infty$.

We also assume the BS has imperfect knowledge of the instantaneous channel realizations $\{\mathbf{h}_k\}$. As in many other works [15], [24], [25], this is modeled by the generic Gauss-Markov formulation form $\forall k$:

$$\widehat{\mathbf{h}}_k = \sqrt{\beta_k}(\sqrt{1-\eta^2}\mathbf{z}_k + \eta\mathbf{w}_k) \quad (13)$$
$$= \sqrt{1-\eta^2}\mathbf{h}_k + \sqrt{\beta_k}\eta\mathbf{w}_k \quad (14)$$

where $\mathbf{w}_k \sim \mathcal{CN}(\mathbf{0},\mathbf{I}_M)$ accounts for the channel estimation errors independent of the fast fading channel vector $\mathbf{z}_k$. The scalar parameter $\eta \in [0,1]$ indicates the quality of the instantaneous CSI, i.e., $\eta = 0$ corresponds to perfect instantaneous CSI and $\eta = 1$ corresponds to having only statistical channel knowledge.[2] The matrix collecting the estimated channel vectors is denoted $\widehat{\mathbf{H}} = [\widehat{\mathbf{h}}_1,\cdots,\widehat{\mathbf{h}}_K]$.

When only imperfect CSI is available at the BS, the structure of the OLP and OLR is not known (most of the existing solutions in the literature are based on heuristic approaches). To overcome this issue, we assume that the true channels $\{\mathbf{h}_k\}$ are simply replaced by their estimates $\{\widehat{\mathbf{h}}_k\}$ (which is an accurate procedure for good CSI quality). This yields

$$\widehat{\mathbf{g}}_k = \sqrt{\frac{\widehat{p}_k}{K}}\frac{\widehat{\mathbf{v}}_k}{\|\widehat{\mathbf{v}}_k\|} \quad (15)$$

where $\widehat{\mathbf{v}}_k = \left(\sum_{\ell=1,\ell\neq k}^{K}\frac{\widehat{q}_\ell}{K}\widehat{\mathbf{h}}_\ell\widehat{\mathbf{h}}_\ell^H + \frac{1}{\rho}\mathbf{I}_M\right)^{-1}\widehat{\mathbf{h}}_k$ and the coefficients $\{\widehat{q}_k\}$ are obtained as:

$$\widehat{q}_k = \frac{\gamma_k\widehat{\tau}}{\frac{1}{K}\widehat{\mathbf{h}}_k^H\left(\sum_{\ell=1,\ell\neq k}^{K}\frac{\widehat{q}_\ell}{K}\widehat{\mathbf{h}}_\ell\widehat{\mathbf{h}}_\ell^H + \frac{1}{\rho}\mathbf{I}_M\right)^{-1}\widehat{\mathbf{h}}_k} \quad \forall k \quad (16)$$

with

$$\widehat{\tau} = \frac{KP_{\max}}{\sum_{n=1}^{K}\gamma_n\left(\frac{1}{K}\widehat{\mathbf{h}}_n^H\left(\sum_{k=1,k\neq n}^{K}\frac{\widehat{q}_k}{K}\widehat{\mathbf{h}}_k\widehat{\mathbf{h}}_k^H + \frac{1}{\rho}\mathbf{I}_M\right)^{-1}\widehat{\mathbf{h}}_n\right)^{-1}}. \quad (17)$$

The transmit powers $\{\widehat{p}_k\}$ are such that (9) is satisfied after replacing $\{\mathbf{h}_k\}$ with $\{\widehat{\mathbf{h}}_k\}$ and $\tau^\star$ with $\widehat{\tau}$. Next, we resort to the large dimensional analysis and show that $\widehat{q}_k$ and $\widehat{p}_k$ gets asymptotically close to explicit deterministic quantities as $M$ and $K$ grow large as for Assumption 1. These quantities provide some insights on the structure of the precoder and receiver as well as on how the different parameters (such as large scale channel gains, imperfect channel knowledge, UE priorities, maximum transmit power) affect the system performance.

---
[2]Observe that the same $\eta$ is assumed for all UEs only for simplicity. The generalization to different $\eta$'s is straightforward.

## A. Asymptotic analysis of OLP and OLR under imperfect CSI

Our first result is as follows:

**Theorem 1.** *Under the settings of Assumptions 1 and 2, we have that $|\widehat{\tau} - \overline{\tau}| \to 0$ where $\overline{\tau}$ is the unique positive solution to the following fixed point equation:*

$$\overline{\tau} = \frac{\rho P_{\max}}{\frac{1}{K}\sum_{i=1}^{K}\frac{\gamma_i}{\beta_i}} \left( \frac{M}{K} - \frac{1}{K}\sum_{i=1}^{K}\frac{\gamma_i \overline{\tau}}{1+\gamma_i \overline{\tau}} \right). \quad (18)$$

*Also, we have that $\max_k |\widehat{q}_k - \overline{q}_k| \to 0$ where*

$$\overline{q}_k = \frac{\gamma_k}{\beta_k} \frac{P_{\max}}{\frac{1}{K}\sum_{i=1}^{K}\frac{\gamma_i}{\beta_i}}. \quad (19)$$

*Proof:* The proof relies on the observation that all the quantities $\widehat{d}_k \triangleq \frac{\gamma_k}{\beta_k}\frac{\widehat{\tau}}{\widehat{q}_k}$ should converge to the same deterministic limit. Note that to determine this limit, standard tools from random matrix theory cannot be applied since $\{\widehat{d}_k\}$ depends on the channel vectors $\{\widehat{\mathbf{h}}_k\}$ in a non-linear fashion. To overcome this issue, we make use of the techniques developed recently in [26]. Details are provided in Appendix B. ∎

The above theorem provides the explicit form of $\{\overline{q}_k\}$, whose computation requires only knowledge of the UEs priority coefficients $\{\gamma_k\}$ and the channel attenuation coefficients $\{\beta_k\}$. The latter can be easily estimated since they change slowly with time. Observe that in the DL the parameter $q_k^\star$ is known to act as a UE priority parameter that implicitly determines how much interference a specific UE $k$ may induce to the other UEs in the cell [16]. Interestingly, its asymptotic value $\overline{q}_k$ is proportional to $\gamma_k$ and inversely proportional to $\beta_k$. Higher priority is thus given to UEs that require high performance (large $\gamma_k$) and/or have weak propagation conditions (small $\beta_k$). In the UL, $q_k^\star$ corresponds to the transmit power of UE $k$. Consequently, (19) indicates that in the asymptotic regime more power is given to UEs with higher priorities and weaker channel conditions.

The asymptotic transmit powers in DL are given in explicit form as follows:

**Theorem 2.** *Under the settings of Assumptions 1 and 2, we have $\max_k |\widehat{p}_k - \overline{p}_k| \to 0$ where*

$$\overline{p}_k = \frac{\gamma_k}{\beta_k}\frac{\overline{\tau}}{\xi}\left( \frac{\beta_k P_{\max}}{(1+\gamma_k \overline{\tau})^2} + \frac{1}{\rho}\right) \quad (20)$$

*and $\xi$ is positive and given by*

$$\xi = \frac{M}{K} - \frac{1}{K}\sum_{i=1}^{K}\frac{(\gamma_i \overline{\tau})^2}{(1+\gamma_i \overline{\tau})^2}. \quad (21)$$

*Proof:* The proof of the convergence of $\{\widehat{p}_k\}$ follows along the same arguments as those used for $\{\widehat{q}_k\}$, and it is thus omitted for space limitations. ∎

The results of Theorems 1 and 2 can be used to compute an asymptotic expression of the SINRs in DL and UL as provided by the following lemmas:

**Lemma 3.** *Under the settings of Assumptions 1 and 2, we have $\max_k |\text{SINR}_k^{\text{dl}\star} - \overline{\text{SINR}}_k^{\text{dl}}| \to 0$ where*

$$\overline{\text{SINR}}_k^{\text{dl}} = \frac{\overline{p}_k(1-\eta^2)\xi}{\mu_k P_{\max} + \frac{1}{\rho \beta_k}} \quad (22)$$

*with*

$$\mu_k = \frac{1 + 2\eta^2 \gamma_k \overline{\tau} + \eta^2 \gamma_k^2 \overline{\tau}^2}{(1+\gamma_k \overline{\tau})^2} \quad (23)$$

*Proof:* By using standard calculus from random matrix theory, it is easily seen that the asymptotic expression of $\{\text{SINR}_k^{\text{dl}}\}$ remains almost surely the same if $\widehat{p}_k$ and $\widehat{q}_k$ are replaced by $\overline{p}_k$ and $\overline{q}_k$. Then, using similar techniques as those in [13], [17], deterministic equivalents of the signal and interference terms can be computed, leading thus to (22). See Appendix C for details. ∎

**Lemma 4.** *Under the settings of Assumptions 1 and 2, we have $\max_k |\text{SINR}_k^{\text{ul}\star} - \overline{\text{SINR}}_k^{\text{ul}}| \to 0$ where*

$$\overline{\text{SINR}}_k^{\text{ul}} = \frac{\overline{q}_k(1-\eta^2)\xi}{\frac{1}{K}\sum_{i=1}^{K}\frac{\beta_i}{\beta_k}\mu_i \overline{q}_i + \frac{1}{\rho \beta_k}}. \quad (24)$$

*Proof:* The proof relies on the same techniques used in Appendix C and it is thus omitted. ∎

An important consequence of the above results is that the performance of the network in DL and UL remains asymptotically the same if $\{\widehat{q}_k\}$ and $\{\widehat{p}_k\}$ are replaced with $\{\overline{q}_k\}$ and $\{\overline{p}_k\}$ such that the precoding/receiving vector of UE $k$ is computed as:

$$\overline{\mathbf{g}}_k = \sqrt{\frac{\overline{p}_k}{K}}\frac{\overline{\mathbf{v}}_k}{\|\overline{\mathbf{v}}_k\|} \quad (25)$$

with

$$\overline{\mathbf{v}}_k = \left( \sum_{i=1}^{K}\frac{\overline{q}_i}{K}\widehat{\mathbf{h}}_i\widehat{\mathbf{h}}_i^H + \frac{1}{\rho}\mathbf{I}_M \right)^{-1}\widehat{\mathbf{h}}_k. \quad (26)$$

This result is particularly interesting from an implementation point of view. Indeed, unlike $\{\widehat{q}_k\}$ and $\{\widehat{p}_k\}$, $\{\overline{q}_k\}$ and $\{\overline{p}_k\}$ in (19) and (20) depend only on the large-scale channel statistics. As a consequence, $\{\overline{q}_k\}$ and $\{\overline{p}_k\}$ are not required to be computed at every channel realization but only once per coherence period. This provides a substantial reduction in computational complexity as compared to OLP and OLR since solving (5) and (6) at the pace of fast fading channel is no longer required. The above analysis reveals also that the asymptotic SINRs in (22) and (24) are both decreasing functions of $\eta$ with maximal value achieved for $\eta = 0$ and given by

$$\frac{\overline{\text{SINR}}_{k,\max}^{\text{dl}}}{\gamma_k} = \frac{\overline{\text{SINR}}_{k,\max}^{\text{ul}}}{\gamma_k} = \overline{\tau} \quad (27)$$

from which it follows that:

**Corollary 1.** *If perfect CSI is available, then in the asymptotic regime the minimum weighted SINR is the same for both DL and UL.*

Unlike the SINR expressions, the coefficients $\overline{q}_k$ and $\overline{p}_k$ in (19) and (20) are found to be independent of $\eta$. This is due to the fact that they depend solely on the statistics of estimated channel vectors $\widehat{\mathbf{h}}_k$, which are the same of the true channel vectors $\mathbf{h}_k$ regardless of the value of $\eta$.[3] Next, we follow a different approach, which aims at designing the OLP and OLR by exploiting the above large system analysis. As shown next, the idea is to still use the vectors $\overline{\mathbf{v}}_k$ in (26) but to design the DL and UL transmit powers so as to maximize the asymptotic minimum SINR.

### B. Asymptotic design of OLP and OLR with imperfect CSI

To begin with, let us call $\tilde{\mathbf{p}} = [\tilde{p}_1, \ldots, \tilde{p}_K]^T$ and $\tilde{\mathbf{q}} = [\tilde{q}_1, \ldots, \tilde{q}_K]^T$ the DL and UL power vectors, respectively, and assume that they are kept fixed. Assume also that the precoding vectors are computed as

$$\tilde{\mathbf{g}}_k = \sqrt{\frac{\tilde{p}_k}{K}} \frac{\overline{\mathbf{v}}_k}{\|\overline{\mathbf{v}}_k\|} \tag{28}$$

with $\overline{\mathbf{v}}_k$ given by (26). Therefore, a direct application of Lemma 3 yields $\max_k |\mathrm{SINR}_k^{\mathrm{dl}}(\tilde{\mathbf{p}}) - \widetilde{\mathrm{SINR}}_k^{\mathrm{dl}}(\tilde{\mathbf{p}})| \to 0$ with

$$\widetilde{\mathrm{SINR}}_k^{\mathrm{dl}}(\tilde{\mathbf{p}}) = \frac{\tilde{p}_k(1-\eta^2)\xi}{\frac{\mu_k}{K}\sum_{i=1}^{K}\tilde{p}_i + \frac{1}{\rho\beta_k}}. \tag{29}$$

Accordingly, from Lemma 4 it follows that $\max_k |\mathrm{SINR}_k^{\mathrm{ul}}(\tilde{\mathbf{q}}) - \widetilde{\mathrm{SINR}}_k^{\mathrm{ul}}(\tilde{\mathbf{q}})| \to 0$ where

$$\widetilde{\mathrm{SINR}}_k^{\mathrm{ul}}(\tilde{\mathbf{q}}) = \frac{\tilde{q}_k(1-\eta^2)\xi}{\frac{1}{K}\sum_{i=1}^{K}\frac{\beta_i}{\beta_k}\mu_i\tilde{q}_i + \frac{1}{\rho\beta_k}}. \tag{30}$$

The main contribution of this section unfolds from the above results and provides the DL and UL power vectors $\tilde{\mathbf{p}}$ and $\tilde{\mathbf{q}}$ that maximize the asymptotic minimum SINR in DL and UL under the power constraint $P_{\max}$. This amounts to solving the following optimization problems:

$$\mathcal{P}_{\mathrm{dl}}^{\mathcal{A}} : \begin{cases} \max_{\tilde{\mathbf{p}}} & \min_k \frac{\widetilde{\mathrm{SINR}}_k^{\mathrm{dl}}(\tilde{\mathbf{p}})}{\gamma_k} \\ \mathrm{s.t.} & \frac{1}{K}\mathbf{1}_K^T\tilde{\mathbf{p}} \le P_{\max} \end{cases} \tag{31}$$

and

$$\mathcal{P}_{\mathrm{ul}}^{\mathcal{A}} : \begin{cases} \max_{\tilde{\mathbf{q}}} & \min_k \frac{\widetilde{\mathrm{SINR}}_k^{\mathrm{ul}}(\tilde{\mathbf{q}})}{\gamma_k} \\ \mathrm{s.t.} & \frac{1}{K}\mathbf{1}_K^T\tilde{\mathbf{q}} \le P_{\max}. \end{cases} \tag{32}$$

Define the diagonal matrix $\mathbf{D} \in \mathbb{C}^{K \times K}$

$$\mathbf{D} = \mathrm{diag}\left(\frac{\gamma_1}{\xi\beta_1(1-\eta^2)}, \cdots, \frac{\gamma_K}{\xi\beta_K(1-\eta^2)}\right) \tag{33}$$

and the vector $\mathbf{f} \in \mathbb{C}^K$ with entries given by $[\mathbf{f}]_i = \beta_i\mu_i/K$. Using the above notation, $\mathcal{P}_{\mathrm{dl}}^{\mathcal{A}}$ and $\mathcal{P}_{\mathrm{ul}}^{\mathcal{A}}$ can be rewritten as

$$\mathcal{P}_{\mathrm{dl}}^{\mathcal{A}} : \begin{cases} \max_{\tilde{\mathbf{p}}} & \min_k \frac{\tilde{p}_k}{\left[\mathbf{D}(\mathbf{f}\mathbf{1}^T\tilde{\mathbf{p}} + \frac{1}{\rho}\mathbf{1})\right]_k} \\ \mathrm{s.t.} & \frac{1}{K}\mathbf{1}_K^T\tilde{\mathbf{p}} \le P_{\max} \end{cases} \tag{34}$$

[3]Observe that $\mathrm{E}\{\widehat{\mathbf{h}}_k\} = \mathrm{E}\{\mathbf{h}_k\} = \mathbf{0}$, $\mathrm{E}\{\widehat{\mathbf{h}}_k\widehat{\mathbf{h}}_k^H\} = \mathrm{E}\{\mathbf{h}_k\mathbf{h}_k^H\} = \beta_k\mathbf{I}_M$ and $\mathrm{E}\{\widehat{\mathbf{h}}_i\widehat{\mathbf{h}}_k^H\} = \mathrm{E}\{\mathbf{h}_i\mathbf{h}_k^H\} = \mathbf{0}_M$.

and

$$\mathcal{P}_{\mathrm{ul}}^{\mathcal{A}} : \begin{cases} \max_{\tilde{\mathbf{q}}} & \min_k \frac{\tilde{q}_k}{\left[\mathbf{D}(\mathbf{1}\mathbf{f}^T\tilde{\mathbf{q}} + \frac{1}{\rho}\mathbf{1})\right]_k} \\ \mathrm{s.t.} & \frac{1}{K}\mathbf{1}_K^T\tilde{\mathbf{q}} \le P_{\max} \end{cases} \tag{35}$$

Following [9], it can be easily shown that $\mathcal{P}_{\mathrm{dl}}^{\mathcal{A}}$ and $\mathcal{P}_{\mathrm{ul}}^{\mathcal{A}}$ are related by the UL-DL duality. Therefore, from [9], [10] it follows that the optimal power vectors $\tilde{\mathbf{p}}^\star$ and $\tilde{\mathbf{q}}^\star$ are such that:

$$\tilde{\mathbf{p}}^\star \propto \mathbf{D}\left(\mathbf{f}\mathbf{1}^T + \frac{1}{\rho K P_{\max}}\mathbf{1}\mathbf{1}^T\right)\tilde{\mathbf{p}}^\star \tag{36}$$

and

$$\tilde{\mathbf{q}}^\star \propto \mathbf{D}\left(\mathbf{1}\mathbf{f}^T + \frac{1}{\rho K P_{\max}}\mathbf{1}\mathbf{1}^T\right)\tilde{\mathbf{q}}^\star. \tag{37}$$

As seen, $\tilde{\mathbf{p}}^\star$ and $\tilde{\mathbf{q}}^\star$ are proportional to the Perron eigenvectors [27] of the non negative matrices $\mathbf{D}(\mathbf{f}\mathbf{1}^T + \frac{1}{\rho K P_{\max}}\mathbf{1}\mathbf{1}^T)$ and $\mathbf{D}(\mathbf{1}\mathbf{f}^T + \frac{1}{\rho K P_{\max}}\mathbf{1}\mathbf{1}^T)$ respectively. Using the inequality constraints, we finally obtain:

$$\tilde{\mathbf{p}}^\star = \frac{K P_{\max}}{\mathbf{1}^T \mathbf{D}(\mathbf{f} + \frac{1}{\rho K P_{\max}}\mathbf{1})}\mathbf{D}(\mathbf{f} + \frac{1}{\rho K P_{\max}}\mathbf{1}) \tag{38}$$

and

$$\tilde{\mathbf{q}}^\star = \frac{K P_{\max}}{\mathbf{1}^T \mathbf{D}\mathbf{1}}\mathbf{D}\mathbf{1}. \tag{39}$$

In a more explicit form, we have that:

$$\tilde{p}_k^\star = \frac{P_{\max}\gamma_k\mu_k + \frac{\gamma_k}{\rho\beta_k}}{\frac{1}{K}\sum_{i=1}^{K}\gamma_i\mu_i + \frac{\gamma_i}{\rho\beta_i P_{\max}}} \quad \text{and} \quad \tilde{q}_k^\star = \frac{\gamma_k}{\beta_k}\frac{P_{\max}}{\frac{1}{K}\sum_{i=1}^{K}\frac{\gamma_i}{\beta_i}}. \tag{40}$$

Unlike $\{\overline{p}_k\}$ in (20), the DL powers $\{\tilde{p}_k^\star\}$ depend on the channel estimation accuracy through $\{\mu_i\}$. This makes the so-called asymptotic OLP (A-OLP) achieve better performance than OLP as shown later by simulations. On the other hand, the UL powers $\{\tilde{q}_k^\star\}$ coincide with $\{\overline{q}_k\}$ in (19), obtained by computing the deterministic equivalents of $\{\widehat{q}_k\}$. This is due to the fact that both solutions rely on the same beamforming receive directions $\widehat{\mathbf{v}}_k$. Therefore, the asymptotic OLR (A-OLR) is identical to OLR.

As mentioned before for OLP and OLR, the use of $\{\tilde{q}_k^\star\}$ and $\{\tilde{p}_k^\star\}$ largely simplifies the implementation of A-OLP and A-OLR as their computation requires only knowledge of the large scale channel statistics and must be performed only once per coherence period (rather than at the same pace as the small-scale fading). Despite being simplified, the implementation of A-OLP and A-OLR as well as that of OLP and OLR still requires the matrix inversion operation in (26). This can be a task of a prohibitively high complexity when $M$ and $K$ are large as envisioned in large scale MIMO systems. To address this issue, a TPE approach will be adopted next.

## IV. USER SPECIFIC TPE PRECODING AND RECEIVER

The common way to apply the TPE concept consists in replacing the matrix inverse by a weighted matrix polynomial with $J$ terms [17], [28]. Differently from the traditional approach, we propose in this work to apply the truncation artifice separately to each vector of the precoding and receiving matrices.

Applying the TPE on a per-UE basis, the precoding vector associated with UE $k$ writes as:

$$\mathbf{g}_{k,\text{TPE}}^{\text{dl}} = \sqrt{\frac{p_{k,\text{TPE}}}{K}} \frac{\mathbf{v}_{k,\text{TPE}}}{\|\mathbf{v}_{k,\text{TPE}}\|} \quad (41)$$

with

$$\mathbf{v}_{k,\text{TPE}} = \sum_{\ell=0}^{J-1} w_{\ell,k}^{\text{dl}} \left(\frac{\widehat{\mathbf{H}}\overline{\mathbf{Q}}\widehat{\mathbf{H}}^H}{K}\right)^\ell \frac{\widehat{\mathbf{h}}_k}{\sqrt{K}} \quad (42)$$

where $J$ is the truncation order and $\overline{\mathbf{Q}} = \text{diag}(\overline{q}_1, \cdots, \overline{q}_K)$. Plugging $\mathbf{g}_{k,\text{TPE}}^{\text{dl}}$ into (2) and letting $\mathbf{w}_{k,\text{dl}} = [w_{0,k}^{\text{dl}}, \cdots, w_{J-1,k}^{\text{dl}}]^T$, the SINR corresponding to UE $k$ can be written as:

$$\text{SINR}_{k,\text{TPE}}^{\text{dl}} = \frac{p_{k,\text{TPE}} \frac{\mathbf{w}_{k,\text{dl}}^H \mathbf{a}_k \mathbf{a}_k^H \mathbf{w}_{k,\text{dl}}}{\mathbf{w}_{k,\text{dl}}^H \mathbf{E}_k \mathbf{w}_{k,\text{dl}}}}{\sum_{i \neq k} \frac{p_{i,\text{TPE}}}{K} \frac{\mathbf{w}_{i,\text{dl}}^H \mathbf{B}_{k,i} \mathbf{w}_{i,\text{dl}}}{\mathbf{w}_{i,\text{dl}}^H \mathbf{E}_i \mathbf{w}_{i,\text{dl}}} + \frac{1}{\rho}} \quad (43)$$

where $\mathbf{a}_k \in \mathbb{C}^{J \times 1}$, $\mathbf{b}_k \in \mathbb{C}^{J \times 1}$, and $\mathbf{B}_{i,k} \in \mathbb{C}^{J \times J}$ are given by:

$$[\mathbf{a}_k]_\ell = \frac{1}{K} \mathbf{h}_k^H \left(\frac{\widehat{\mathbf{H}}\overline{\mathbf{Q}}\widehat{\mathbf{H}}^H}{K}\right)^\ell \widehat{\mathbf{h}}_k \quad (44)$$

$$[\mathbf{B}_{k,i}]_{\ell,m} = \frac{1}{K} \mathbf{h}_k^H \left(\frac{\widehat{\mathbf{H}}\overline{\mathbf{Q}}\widehat{\mathbf{H}}^H}{K}\right)^\ell \widehat{\mathbf{h}}_i \widehat{\mathbf{h}}_i^H \left(\frac{\widehat{\mathbf{H}}\overline{\mathbf{Q}}\widehat{\mathbf{H}}^H}{K}\right)^m \mathbf{h}_k \quad (45)$$

$$[\mathbf{E}_k]_{\ell,m} = \frac{1}{K} \widehat{\mathbf{h}}_k^H \left(\frac{\widehat{\mathbf{H}}\overline{\mathbf{Q}}\widehat{\mathbf{H}}^H}{K}\right)^{\ell+m} \widehat{\mathbf{h}}_k. \quad (46)$$

The transmit power at the BS can be easily found as $P = \frac{1}{K} \sum_{k=1}^{K} p_{k,\text{TPE}}$.

The TPE concept is now applied in UL to the OLR. Let $\left\{\frac{q_{k,\text{TPE}}}{K}\right\}$ be the set of UL transmit powers. The receive beamforming vector associated with UE $k$ is thus given by:

$$\mathbf{g}_{k,\text{TPE}}^{\text{ul}} = \sum_{\ell=0}^{J-1} w_{\ell,k}^{\text{ul}} \left(\frac{\widehat{\mathbf{H}}\overline{\mathbf{Q}}\widehat{\mathbf{H}}^H}{K}\right)^\ell \frac{\widehat{\mathbf{h}}_k}{\sqrt{K}}. \quad (47)$$

Plugging $\mathbf{g}_{k,\text{TPE}}^{\text{ul}}$ into (12) yields the SINR of UE $k$ given by:

$$\text{SINR}_{k,\text{TPE}}^{\text{ul}} = \frac{q_{k,\text{TPE}} \mathbf{w}_{k,\text{ul}}^H \mathbf{a}_k \mathbf{a}_k^H \mathbf{w}_{k,\text{ul}}}{\sum_{i \neq k} \frac{q_{i,\text{TPE}}}{K} \mathbf{w}_{k,\text{ul}}^H \mathbf{B}_{i,k} \mathbf{w}_{k,\text{ul}} + \frac{1}{\rho} \mathbf{w}_{k,\text{ul}}^H \mathbf{E}_k \mathbf{w}_{k,\text{ul}}} \quad (48)$$

where $\mathbf{w}_{k,\text{ul}} = \left[w_{0,k}^{\text{ul}}, \cdots, w_{J-1,k}^{\text{ul}}\right]^T$ and $\mathbf{a}_k$, $\mathbf{B}_{i,k}$ $\mathbf{E}_k$ are given by (44) – (46).

## V. ASYMPTOTIC ANALYSIS AND OPTIMIZATION OF THE USER SPECIFIC TPE PRECODER AND RECEIVER

In this section, we consider the asymptotic regime defined in Assumption 1 and show that the SINRs of the TPE precoder and receiver converge to deterministic equivalents, that depend only on the weighting vectors $\{\mathbf{w}_{k,\text{dl}}\}_{k=1}^{K}$ or $\{\mathbf{w}_{k,\text{ul}}\}_{k=1}^{K}$, and the large-scale channel statistics. These deterministic equivalents are then exploited to compute the optimal weights that maximize the minimum asymptotic DL/UL SINR.

### A. Asymptotic Analysis

Let us introduce the fundamental equations that are needed to express the deterministic equivalents. We begin by defining $\delta(t)$ as the unique positive solution of the following equation $\forall t > 0$:

$$\delta(t) = \frac{M}{K} \frac{1}{1 + \frac{t}{K} \sum_{i=1}^{K} \frac{\beta_i \overline{q}_i}{1 + t \delta(t) \overline{q}_i \beta_i}}. \quad (50)$$

Define $\overline{X}_k(t)$ and $\overline{Z}_{k,i}(t)$ as:

$$\overline{X}_k(t) = \frac{\beta_k \delta(t)}{1 + t \overline{q}_k \beta_k \delta(t)}, \quad (51)$$

$$\overline{Z}_{k,i}(t,u) = \frac{\beta_i \overline{f}_k(t,u) \overline{\alpha}(t,u)}{(1 + t \delta(t) \beta_i \overline{q}_i)(1 + u \delta(u) \beta_i \overline{q}_i)}, \quad (52)$$

with $\overline{f}_k(t,u)$ being given by:

$$\overline{f}_k(t,u) = \beta_k \left( \eta^2 + \frac{1 - \eta^2}{(1 + \overline{q}_k \beta_k t \delta(t))(1 + \overline{q}_k \beta_k u \delta(u))} \right) \quad (53)$$

and

$$\overline{\alpha}(t,u) = \frac{\delta(t)\delta(u)}{\frac{M}{K} - \frac{tu}{K} \delta(t) \delta(u) \sum_{i=1}^{K} \frac{[\beta_i \overline{q}_i]^2}{[1 + t \overline{q}_i \beta_i \delta(t)][1 + u \overline{q}_i \beta_i \delta(u)]}}. \quad (54)$$

Let $\overline{\mathbf{a}}_k \in \mathbb{C}^J$ be defined as:

$$[\overline{\mathbf{a}}_k]_\ell = \frac{(-1)^\ell}{\ell!} \sqrt{1 - \tau^2} \overline{X}_k^{(\ell)}, \quad (55)$$

and call $\overline{\mathbf{B}}_{i,k} \in \mathbb{C}^{J \times J}$ and $\overline{\mathbf{E}}_k \in \mathbb{C}^{J \times J}$ the matrices with elements given by:

$$\left[\overline{\mathbf{B}}_{k,i}\right]_{\ell,m} = \frac{(-1)^{\ell+m}}{\ell!m!} \overline{Z}_{k,i}^{(\ell+m)} \quad (56)$$

$$\left[\overline{\mathbf{E}}_k\right]_{\ell,m} = \frac{(-1)^{\ell+m}}{(\ell+m)!} \overline{X}_k^{(\ell+m)}. \quad (57)$$

The main technical result of this section then lies in the following lemma:

**Lemma 5.** *Under the settings of Assumptions 1 and 2, we have* $\max_k |\text{SINR}_{k,\text{TPE}}^{\text{dl}} - \overline{\text{SINR}}_{k,\text{TPE}}^{\text{dl}}| \to 0$ *and* $\max_k |\text{SINR}_{k,\text{TPE}}^{\text{ul}} - \overline{\text{SINR}}_{k,\text{TPE}}^{\text{ul}}| \to 0$ *with*

$$\overline{\text{SINR}}_{k,\text{TPE}}^{\text{dl}} = \frac{p_{k,\text{TPE}} \frac{\mathbf{w}_{k,\text{dl}}^H \overline{\mathbf{a}}_k \overline{\mathbf{a}}_k^H \mathbf{w}_{k,\text{dl}}}{\mathbf{w}_{k,\text{dl}}^H \overline{\mathbf{E}}_k \mathbf{w}_{k,\text{dl}}}}{\sum_{i \neq k} \frac{p_{i,\text{TPE}}}{K} \frac{\mathbf{w}_{i,\text{dl}}^H \overline{\mathbf{B}}_{k,i} \mathbf{w}_{i,\text{dl}}}{\mathbf{w}_{i,\text{dl}}^H \overline{\mathbf{E}}_k \mathbf{w}_{i,\text{dl}}} + \frac{1}{\rho}} \quad (58)$$

$$\overline{\text{SINR}}_{k,\text{TPE}}^{\text{ul}} = \frac{q_{k,\text{TPE}} \mathbf{w}_{k,\text{ul}}^H \overline{\mathbf{a}}_k \overline{\mathbf{a}}_k^H \mathbf{w}_{k,\text{ul}}}{\sum_{i \neq k} \frac{q_{i,\text{TPE}}}{K} \mathbf{w}_{k,\text{ul}}^H \overline{\mathbf{B}}_{i,k} \mathbf{w}_{k,\text{ul}} + \frac{1}{\rho} \mathbf{w}_{k,\text{ul}}^H \overline{\mathbf{E}}_k \mathbf{w}_{k,\text{ul}}}. \quad (59)$$

*Also, we have that* $P - \overline{P} \to 0$ *with*

$$\overline{P} = \frac{1}{K} \sum_{k=1}^{K} \mathbf{w}_k^H \overline{\mathbf{E}}_k \mathbf{w}_k. \quad (60)$$

*Proof:* The deterministic equivalents of the SINR and transmit powers are obtained by computing the asymptotic

$$q_{k,\text{TPE}}^\star = \frac{\gamma_k K P_{\max}}{\sum_{\ell=1}^{K} \frac{\gamma_\ell \overline{\mathbf{a}}_k^T \overline{\mathbf{E}}_k^{-\frac{1}{2}} \left( \sum_{i \neq k} \frac{\rho}{K} q_{i,\text{TPE}}^\star \overline{\mathbf{E}}_k^{-\frac{1}{2}} \overline{\mathbf{B}}_{i,k} \overline{\mathbf{E}}_k^{-\frac{1}{2}} + \mathbf{I}_J \right)^{-1} \overline{\mathbf{E}}_k^{-\frac{1}{2}} \overline{\mathbf{a}}_k}{\overline{\mathbf{a}}_\ell^T \overline{\mathbf{E}}_\ell^{-\frac{1}{2}} \left( \sum_{j \neq \ell} \frac{\rho}{K} q_{j,\text{TPE}}^\star \overline{\mathbf{E}}_\ell^{-\frac{1}{2}} \overline{\mathbf{B}}_{j,\ell} \overline{\mathbf{E}}_\ell^{-\frac{1}{2}} + \mathbf{I}_J \right)^{-1} \overline{\mathbf{E}}_\ell^{-\frac{1}{2}} \overline{\mathbf{a}}_\ell}} \quad \forall k \quad (49)$$

---

expressions of the entries of $\mathbf{a}_k$, $\mathbf{B}_{k,i}$ and $\mathbf{E}_k$. The latter can be written as a function of the derivatives of some quadratic forms whose deterministic equivalents are known in random matrix theory. See Appendix D for details. ∎

With the asymptotic equivalents of the SINR and the transmit power on hand, we are ready now to determine the optimal parameters of the TPE based receiver and precoder.

### B. Optimization of the US-TPE precoding and receiver

In the sequel, we compute the optimal weighting vectors $\mathbf{w}_{k,\text{dl}}$ and $\mathbf{w}_{k,\text{ul}}$ as well as the optimal DL and UL transmit powers. To begin with, we let

$$\mathbf{c}_{k,\text{dl}} = \frac{\overline{\mathbf{E}}_k^{\frac{1}{2}} \mathbf{w}_{k,\text{dl}}}{\|\overline{\mathbf{E}}_k^{\frac{1}{2}} \mathbf{w}_{k,\text{dl}}\|} \quad \mathbf{c}_{k,\text{ul}} = \frac{\overline{\mathbf{E}}_k^{\frac{1}{2}} \mathbf{w}_{k,\text{ul}}}{\|\overline{\mathbf{E}}_k^{\frac{1}{2}} \mathbf{w}_{k,\text{ul}}\|} \quad (61)$$

and rewrite the asymptotic SINR expressions in (58) and (59) as follows:

$$\overline{\text{SINR}}_{k,\text{TPE}}^{\text{dl}} = \frac{p_{k,\text{TPE}} \, \mathbf{c}_{k,\text{dl}}^H \overline{\mathbf{E}}_k^{-\frac{1}{2}} \overline{\mathbf{a}}_k \overline{\mathbf{a}}_k^H \overline{\mathbf{E}}_k^{-\frac{1}{2}} \mathbf{c}_{k,\text{dl}}}{\sum_{i \neq k} \frac{p_{i,\text{TPE}}}{K} \, \mathbf{c}_{i,\text{dl}}^H \overline{\mathbf{E}}_i^{-\frac{1}{2}} \overline{\mathbf{B}}_{k,i} \overline{\mathbf{E}}_i^{-\frac{1}{2}} \mathbf{c}_{i,\text{dl}} + \frac{1}{\rho}} \quad (62)$$

$$\overline{\text{SINR}}_{k,\text{TPE}}^{\text{ul}} = \frac{q_{k,\text{TPE}} \mathbf{c}_{k,\text{ul}}^H \overline{\mathbf{E}}_k^{-\frac{1}{2}} \overline{\mathbf{a}}_k \overline{\mathbf{a}}_k^H \overline{\mathbf{E}}_k^{-\frac{1}{2}} \mathbf{c}_{k,\text{ul}}}{\sum_{i \neq k} \frac{q_{i,\text{TPE}}}{K} \mathbf{c}_{k,\text{ul}}^H \overline{\mathbf{E}}_k^{-\frac{1}{2}} \overline{\mathbf{B}}_{i,k} \overline{\mathbf{E}}_k^{-\frac{1}{2}} \mathbf{c}_{k,\text{ul}} + \frac{1}{\rho}}. \quad (63)$$

The parameters $\{\mathbf{c}_{k,\text{dl}}\}$, $\{\mathbf{c}_{k,\text{ul}}\}$, $\{p_{k,\text{TPE}}\}$ and $\{q_{k,\text{TPE}}\}$ are computed as solutions of the following optimization problems:

$$\mathcal{P}_{\text{dl}}^{\text{TPE}} : \begin{cases} \max_{\{\mathbf{c}_{k,\text{dl}}\}, \mathbf{p}_{\text{TPE}}} \min_k \frac{\overline{\text{SINR}}_{k,\text{TPE}}^{\text{dl}}}{\gamma_k} \\ \text{s.t.} \quad \frac{1}{K} \mathbf{1}_K^T \mathbf{p}_{\text{TPE}} \leq P_{\max} \end{cases} \quad (64)$$

and

$$\mathcal{P}_{\text{ul}}^{\text{TPE}} : \begin{cases} \max_{\{\mathbf{c}_{k,\text{ul}}\}, \mathbf{q}_{\text{TPE}}} \min_k \frac{\overline{\text{SINR}}_{k,\text{TPE}}^{\text{ul}}}{\gamma_k} \\ \text{s.t.} \quad \frac{1}{K} \mathbf{1}_K^T \mathbf{q}_{\text{TPE}} \leq P_{\max} \end{cases} \quad (65)$$

which have the same structure of (3) and (11). Following similar arguments, it turns out that the solution is such that all the weighted asymptotic SINRs are equal to $\tau_{\text{TPE}}^\star$:

$$\tau_{\text{TPE}}^\star = \frac{\overline{\text{SINR}}_{k,\text{TPE}}^{\text{ul}}}{\gamma_k} = \frac{\overline{\text{SINR}}_{k,\text{TPE}}^{\text{dl}}}{\gamma_k} \quad \forall k. \quad (66)$$

The optimal values $q_{k,\text{TPE}}^\star$ are obtained as the unique solution of the fixed-point system of equations in (49) whereas the optimal weighting vectors are such that $\mathbf{c}_{k,\text{ul}}^\star = \mathbf{c}_{k,\text{dl}}^\star = \mathbf{c}_k^\star$ with:

$$\mathbf{c}_k^\star = \frac{\left( \sum_{i \neq k} \frac{q_{i,\text{TPE}}^\star}{K} \overline{\mathbf{E}}_k^{-\frac{1}{2}} \overline{\mathbf{B}}_{i,k} \overline{\mathbf{E}}_k^{-\frac{1}{2}} + \frac{1}{\rho} \mathbf{I}_J \right)^{-1} \overline{\mathbf{E}}_k^{-\frac{1}{2}} \overline{\mathbf{a}}_k}{\left\| \left( \sum_{i \neq k} \frac{q_{i,\text{TPE}}^\star}{K} \overline{\mathbf{E}}_k^{-\frac{1}{2}} \overline{\mathbf{B}}_{i,k} \overline{\mathbf{E}}_k^{-\frac{1}{2}} + \frac{1}{\rho} \mathbf{I}_J \right)^{-1} \overline{\mathbf{E}}_k^{-\frac{1}{2}} \overline{\mathbf{a}}_k \right\|}. \quad (67)$$

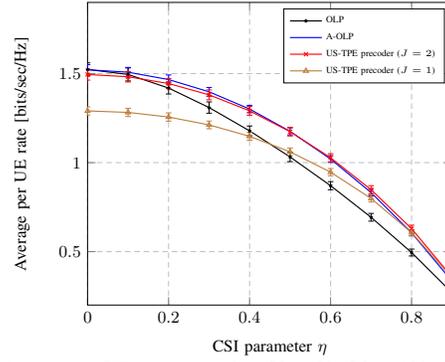

Fig. 1. Average per UE rate vs. $\eta$ when $K = 32$, $M = 128$, $P_{\max} = 5$ Watt and $\rho = 20$ dB.

From (49), it follows that the computation of $\{q_{k,\text{TPE}}^\star\}$ requires matrix inversions whose complexity depend on $J$. However, $J$ is small and does not need to scale with the values of $M$ and $K$. Thus, the computation of $q_{k,\text{TPE}}^\star$ is not very demanding. The optimal power vector $\mathbf{p}_{\text{TPE}}^\star$ is such that the weighted SINRs in the uplink are all equal to $\tau_{\text{TPE}}^\star$. This yields:

$$\mathbf{p}_{\text{TPE}} = \frac{\tau_{\text{TPE}}^\star}{\rho} \left( \mathbf{I}_K - \tau_{\text{TPE}}^\star \mathbf{\Gamma}_{\text{TPE}} \mathbf{F}_{\text{TPE}} \right)^{-1} \mathbf{\Gamma}_{\text{TPE}} \mathbf{1}_K \quad (68)$$

where:

$$\mathbf{\Gamma}_{\text{TPE}} = \text{diag} \left\{ \left( \mathbf{c}_{k,\text{dl}}^H \overline{\mathbf{E}}_k^{-\frac{1}{2}} \overline{\mathbf{a}}_k \overline{\mathbf{a}}_k^H \overline{\mathbf{E}}_k^{-\frac{1}{2}} \mathbf{c}_{k,\text{dl}} \right)^{-1} \right\}_{k=1}^{K} \quad (69)$$

and

$$[\mathbf{F}_{\text{TPE}}]_{k,i} = \begin{cases} 0 & \text{if } k = i \\ \frac{1}{K} \mathbf{c}_{i,\text{dl}}^H \overline{\mathbf{E}}_k^{-\frac{1}{2}} \overline{\mathbf{B}}_{k,i} \overline{\mathbf{E}}_k^{-\frac{1}{2}} \mathbf{c}_{i,\text{dl}} & \text{if } k \neq i. \end{cases} \quad (70)$$

From the above results, it follows that the TPE-based schemes have the same structure as A-OLP and A-OLR. However, the former allow a considerable reduction in the complexity since they require only about $\mathcal{O}(KM)$ arithmetic operations as they do not involve the computation of a matrix inverse. This has to be compared with the OLP and OLR that involve $\mathcal{O}(K^2 M)$ arithmetic operations. For more details about complexity analysis and saving, we refer the reader to [17] where the benefits of TPE when applied to precoding schemes are discussed in details.

## VI. SIMULATION RESULTS

Numerical results are now used to make comparisons among the different transceiver schemes and to validate the asymptotic analysis. The UEs are assumed to be uniformly distributed in a cell with radius 250 m. The path loss $\beta_k$ between the BS and a UE $k$ with distance $x_k$ from the BS is modeled as $\beta_k = 1/\left[1 + (x_k/d_0)^\delta\right]$ where $\delta = 3.8$, $d_0 = 30$ m. The analysis is conducted in terms of the average achievable rate per UE given by:

$$r = \frac{1}{K} \sum_{k=1}^{K} \mathbb{E}\left[\log_2(1 + \text{SINR}_k)\right] \quad (71)$$

where the expectation is taken with respect to the different channel realization. We set $\rho = 20$ dB and assume that the UEs' priorities $\{\gamma_k\}$ are randomly chosen from the interval $[1, 2]$. Markers are used to represent the asymptotic results whereas the error bars indicate the standard deviation of the Monte Carlo results.

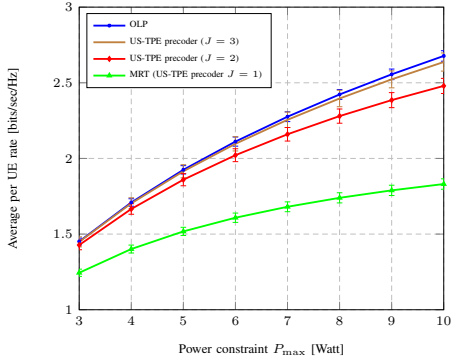

Fig. 2. Average per UE rate vs. power constraint $P_{\max}$ when $K = 32$, $M = 128$, $\rho = 20$ dB and $\eta = 0$.

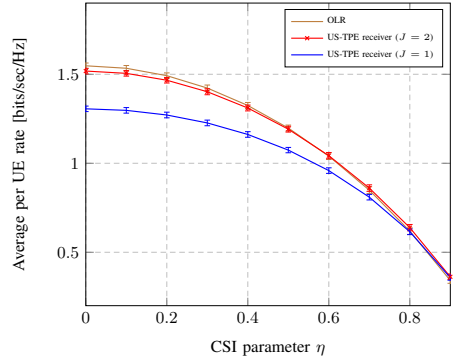

Fig. 3. Average per UE rate vs. $\eta$ when $K = 32$, $M = 128$, $P_{\max} = 5$ Watt and $\rho = 20$ dB.

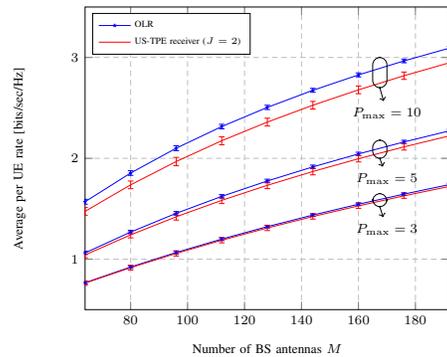

Fig. 4. Average per UE rate vs. $M$ when $K = 32$, $\rho = 20$ dB and $\eta = 0$.

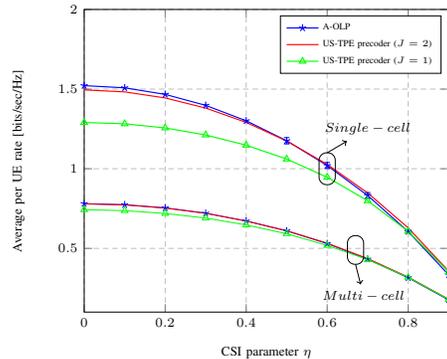

Fig. 5. Average per UE rate vs. $\eta$ when $M = 128$, $K = 32$, $\rho = 20$ dB and $L = 3$ BSs.

Fig. 1 reports the downlink average rate per UE of OLP, A-OLP and US-TPE precoding as a function of $\eta$ when $K = 32$, $M = 128$ and $P_{\max} = 5$ Watt. As seen, when $\eta$ takes small values, A-OLP and OLP have approximately the same performances. As $\eta$ increases, OLP presents a more significant loss in average per UE rate performance. Moreover, it can be seen that US-TPE with $J = 2$ achieves almost the same performance as A-OLP and this over all the range of $\eta$. This clearly confirms that US-TPE shares the same interesting features of A-OLP while requiring a lower complexity.

Fig. 2 investigates the DL average per UE rate with respect to the power budget $P_{\max}$ when $K = 32$, $M = 128$, $\eta = 0$ (perfect CSI case) and $\rho = 20$ dB. An important observation from Fig. 2 is that the gap in performance between US-TPE and OLP increases with $P_{\max}$. To reduce this gap, one solution is to use the US-TPE with higher truncation orders.

A similar analysis is now conducted for the UL. Only OLR is considered since A-OLR and OLR have asymptotically the same performance. Fig. 3 illustrates the uplink average rate per UE vs. $\eta$. As seen, with $J = 2$, US-TPE receiver provides the same performance as OLR.

Fig. 4 illustrates the uplink average rate per UE vs. the number of BS antennas $M$ for different values of $P_{\max}$ when $\eta = 0$. As seen, US-TPE receiver provides comparable performance to OLR for all values of $M$. Besides, the gap increases with $P_{\max}$ as in DL, and seems to be weakly dependent of the number of antennas $M$.

Finally, it is worth pointing out that in a multi-cell scenario, the performance of both the optimal transceivers and our proposed schemes might undergo degradation since their design discards the impact of inter-cell interference. To give flavor of the impact of inter-cell interference, we consider in a last experiment a multi-cell network composed of 3 adjacent cells each of radius $r = 250m$. The BS in every cell is equipped with 128 antennas and communicates with 32 UEs. Fig. 5 displays the average per UE rate vs. the CSI parameter. For sake of comparison, we show in the same figure the results corresponding to the single-cell scenario. As expected, both schemes undergo a degradation in performance caused by discarding the impact of inter-cell interference, but the gap between the two schemes is still negligible.

## VII. CONCLUSIONS

This work considered a single-cell large-scale MU-MIMO system and studied the problem of designing the optimal linear transceivers that maximize the minimum SINR while satisfying a certain power constraint. We considered the asymptotic regime in which the number of BS antennas $M$ and the number of the UEs $K$ grow large with the same pace. Stating and proving new results from large-scale random matrix theory allowed us to give concise approximations of the optimal transceivers. Such approximations turned out to be of much lower complexity as they depend only on the long-term channel attenuations of the UEs, the maximum transmit power and the quality of the channel estimates. Numerical results indicated that these approximations are very accurate even for small system dimensions. To further reduce the computational complexity, we proposed to apply the truncated polynomial expansion technique to the precoding and receiving vectors of each UE. The resulting transceiver was then optimized in the asymptotic regime. Numerical results showed that it achieves a-close-to-optimal performance.

## APPENDIX A
## USEFUL LEMMAS

This appendix gathers some technical results from random matrix theory concerning the asymptotic behaviour of large random matrices. Next, we denote by $\mathbf{X} = [\mathbf{x}_1, \cdots, \mathbf{x}_K]$ a $M \times K$ standard complex Gaussian matrix. Let $t > 0$ and $\mathbf{R} = \mathrm{diag}(\alpha_1, \cdots, \alpha_K)$. We define the resolvent matrix of $\mathbf{XRX}^H$ as:

$$\mathbf{Q}(t) = \left( \frac{t}{K} \sum_{i=1}^{K} \alpha_i \mathbf{x}_i \mathbf{x}_i^H + \mathbf{I}_M \right)^{-1} = \left( \frac{t}{K} \mathbf{XRX}^H + \mathbf{I}_M \right)^{-1}. \tag{72}$$

Define also $\mathbf{Q}_k(t)$ as: $\mathbf{Q}_k(t) = \left( \frac{t}{K} \sum_{i \neq k} \alpha_i \mathbf{x}_i \mathbf{x}_i^H + \mathbf{I}_M \right)^{-1}$ which is obtained from $\mathbf{Q}(t)$ by removing the contribution of vector $\mathbf{x}_k$. The following lemmas recall some classical identities involving the resolvent matrix, which will be extensively used in our derivations:

**Lemma 6.** *The following identities hold true:*

*1) Inverse of resolvents:*

$$\mathbf{Q}(t) = \mathbf{Q}_k(t) - \frac{t\alpha_k \mathbf{Q}_k(t) \mathbf{x}_k \mathbf{x}_k^H \mathbf{Q}_k(t)}{1 + \frac{t\alpha_k}{K} \mathbf{x}_k^H \mathbf{Q}_k(t) \mathbf{x}_k}. \tag{73}$$

*2) Rank-one perturbation result: For any matrix $\mathbf{A}$, we have* $\mathrm{tr}\mathbf{A}\left(\mathbf{Q}(t) - \mathbf{Q}_k(t)\right) \leq \|\mathbf{A}\|_2$.

**Lemma 7** (Convergence of quadratic forms). *Let* $\mathbf{y} \sim \mathcal{CN}(\mathbf{0}_M, \mathbf{I}_M)$. *Let $\mathbf{A}$ be an $M \times M$ matrix independent of $\mathbf{y}$, which has a bounded spectral norm; that is, there exists $C_A < \infty$ such that $\|\mathbf{A}\|_2 \leq C_A$. Then, for any $p \geq 1$, there exists a constant $C_p$ depending only on $p$, such that*

$$\mathbb{E}_\mathbf{y}\left[ \left| \frac{1}{M} \mathbf{y}^H \mathbf{A} \mathbf{y} - \frac{1}{M} \mathrm{tr}(\mathbf{A}) \right|^p \right] \leq \frac{C_p C_A^p}{M^{p/2}}, \tag{74}$$

*By choosing $p \geq 2$, we thus have that*

$$\frac{1}{M} \mathbf{y}^H \mathbf{A} \mathbf{y} - \frac{1}{M} \mathrm{tr}(\mathbf{A}) \to 0. \tag{75}$$

The following lemma provides results allowing to approximate random quantities involving the resolvent matrix when their dimensions grow simultaneously large:

**Lemma 8.** *Let $\delta(t)$ be the unique positive solution to the following equation:*

$$\delta(t) = \frac{M}{K\left(1 + \frac{t}{K} \sum_{i=1}^{K} \frac{\alpha_i}{1 + t\delta(t)\alpha_i}\right)}. \tag{76}$$

*Consider the asymptotic regime in which $M$ and $K$ grow to infinity with: $0 < \liminf \frac{M}{K} < \limsup \frac{M}{K} < \infty$. Let $[a,b]$ be a closed bounded interval in $[0, \infty)$. the following convergences holds true:*

$$\sup_{t \in [a,b]} \left| \frac{1}{K} \mathrm{tr}\, \mathbf{Q}(t) - \delta(t) \right| \to 0. \tag{77}$$

*Moreover, if $\mathbf{y}_1, \cdots, \mathbf{y}_K$ denotes standard complex Gaussian vectors independent from $\mathbf{x}_1, \cdots, \mathbf{x}_K$, we have:*

$$\max_j \sup_{t \in [a,b]} \left| \mathbf{y}_j^H \mathbf{Q}(t) \mathbf{y}_j - \delta(t) \right| \to 0. \tag{78}$$

Note that, as a consequence of the rank-one perturbation lemma, the above convergences can be transferred to the resolvent matrix $\mathbf{Q}_k(t)$. As a matter of fact, we also have:

$$\sup_{t \in [a,b]} \left| \frac{1}{K} \mathrm{tr}\, \mathbf{Q}_k(t) - \delta(t) \right| \to 0. \tag{79}$$

and

$$\max_j \sup_{t \in [a,b]} \left| \mathbf{y}_j^H \mathbf{Q}(t) \mathbf{y}_j - \delta(t) \right| \to 0. \tag{80}$$

## APPENDIX B
## PROOF OF THEOREM 1

We aim at determining deterministic equivalents of $\{\widehat{q}_k\}$. Let $\widehat{\mathbf{z}}_k = \beta_k^{-\frac{1}{2}} \widehat{\mathbf{h}}_k$. Define $\widehat{\mathbf{Q}}_k = (\sum_{m \neq k} \frac{\widehat{q}_\ell}{K} \widehat{\mathbf{h}}_\ell \widehat{\mathbf{h}}_\ell^H + \frac{1}{\rho} \mathbf{I}_M)^{-1}$. Then, $\widehat{q}_k$ writes as: $\widehat{q}_k = \frac{\gamma_k \widehat{\tau}}{\frac{\beta_k}{K} \widehat{\mathbf{z}}_k^H \widehat{\mathbf{Q}}_k \widehat{\mathbf{z}}_k}$. Intuitively, from rank-one perturbation arguments (See Lemma 6), all $\widehat{d}_k \triangleq \frac{1}{K} \widehat{\mathbf{z}}_k^H \widehat{\mathbf{Q}}_k \widehat{\mathbf{z}}_k$ present the same asymptotic behavior and should converge to the same limit. In light of this observation, we will rather focus on the study of the convergence of $\{\widehat{d}_k\}$. The convergence of $\{\widehat{q}_k\}$ to $\{\overline{q}_k\}$ will then follow.

It can be thus easily shown that $\{\widehat{d}_k\}$ are the positive solutions to the following fixed-point equations:

$$\widehat{d}_k = \frac{1}{K} \widehat{\mathbf{z}}_k^H \left( \sum_{m \neq k} \frac{\widehat{\tau} \gamma_m}{K \widehat{d}_m} \widehat{\mathbf{z}}_m \widehat{\mathbf{z}}_m^H + \frac{1}{\rho} \mathbf{I}_M \right)^{-1} \widehat{\mathbf{z}}_k. \tag{81}$$

Note that direct application of standard random matrix theory tools to the quadratic form arising in the expressions of $\{\widehat{d}_k\}$ is not analytically correct since coefficients $\{\widehat{d}_k\}$ and $\widehat{\tau}$ are both function of the channel vectors $\{\mathbf{z}_k\}$. However, one would expect coefficients $\{\widehat{d}_m\}_{m \neq k}$ to be weakly dependent of $\widehat{\mathbf{z}}_k$, and thus considering $\{\widehat{d}_k\}$ as deterministic, although not properly correct, would lead to infer about their asymptotic behavior. Based on these intuitive arguments and using the results of Lemma 7 and Lemma 8 (see (76)) when all $\widehat{d}_k$ are replaced by the same quantity $\widetilde{d}$, one could claim that $\{\widehat{d}_k\}$ must satisfy the following convergence:

$$\max_k \left|\widehat{d}_k/\widetilde{d} - 1\right| \to 0 \quad (82)$$

where $\widetilde{d}$ is given by

$$\frac{\widetilde{d}}{\rho} = \frac{M}{K} - \frac{1}{K}\sum_{m=1}^{K}\frac{\gamma_m \widehat{\tau}}{1+\gamma_m \widehat{\tau}}. \quad (83)$$

It is worth mentioning that $\widetilde{d}$ constitutes an asymptotic random, not deterministic, equivalent of $\widehat{d}$ as it depends on $\widehat{\tau}$. Additional work is needed to find a deterministic equivalent for $\widetilde{d}$. This will be performed later. We will now focus on providing a rigorous proof for (82). To this end, we will make use of the approach developed in [29]. Let us define $e_k = \widehat{d}_k/\widetilde{d}$, and assume without loss of generality that $e_1 < \cdots < e_K$. We can thus write $\{\widehat{d}_k\}$ as:

$$\widehat{d}_k = \frac{1}{K}\widehat{\mathbf{z}}_k^H \left(\sum_{m\neq k}\frac{\widehat{\tau}\gamma_m\widehat{\mathbf{z}}_m\widehat{\mathbf{z}}_m^H}{Ke_m\widetilde{d}} + \frac{1}{\rho}\mathbf{I}_M\right)^{-1}\widehat{\mathbf{z}}_k \quad (84)$$

from which dividing by $\widetilde{d}$ we get:

$$e_k = \frac{1}{K}\widehat{\mathbf{z}}_k^H \left(\sum_{m\neq k}\frac{\widehat{\tau}\gamma_m\widehat{\mathbf{z}}_m\widehat{\mathbf{z}}_m^H}{Ke_m} + \frac{\widetilde{d}}{\rho}\mathbf{I}_M\right)^{-1}\widehat{\mathbf{z}}_k. \quad (85)$$

From monotonicity arguments, it follows that:

$$e_K \le \frac{1}{K}\widehat{\mathbf{z}}_K^H\left(\sum_{m\neq K}\frac{\widehat{\tau}\gamma_m\widehat{\mathbf{z}}_m\widehat{\mathbf{z}}_m^H}{Ke_K} + \frac{\widetilde{d}}{\rho}\mathbf{I}_M\right)^{-1}\widehat{\mathbf{z}}_K \quad (86)$$

or, equivalently,

$$1 \le \frac{1}{K}\widehat{\mathbf{z}}_K^H\left(\sum_{m\neq k}\frac{\widehat{\tau}\gamma_m\widehat{\mathbf{z}}_m\widehat{\mathbf{z}}_m^H}{K} + \frac{\widetilde{d}e_K}{\rho}\mathbf{I}_M\right)^{-1}\widehat{\mathbf{z}}_K. \quad (87)$$

To prove that $\max_k |e_k - 1| \to 0$, we proceed by contradiction. Assume that there exists $\ell > 0$ such that $\limsup e_K > 1 + \ell$. Then, $e_K$ is infinitely often larger than $1+\ell$. Let us restrict ourselves to such a subsequence. Therefore, we have:

$$1 \le \frac{1}{K}\widehat{\mathbf{z}}_K^H\left(\widehat{\tau}\sum_{m\neq k}\frac{\gamma_m\widehat{\mathbf{z}}_m\widehat{\mathbf{z}}_m^H}{K} + \frac{\widetilde{d}(1+\ell)}{\rho}\mathbf{I}_M\right)^{-1}\widehat{\mathbf{z}}_K. \quad (88)$$

We shall now invoke the convergence results of Lemma 8 in Appendix A. But before that, we need to check that $\rho\widehat{\tau}/\widetilde{d}$ stay almost surely in a bounded interval. This can be shown by noticing that function $f$ can be bounded above and below by:

$$\frac{\frac{M}{K}}{\frac{x}{\rho}+1} \le f(x) \le \frac{M\rho}{Kx} \quad (89)$$

and thus:

$$\rho\left(\frac{M}{K}-1\right) \le \widetilde{d} \le \frac{M\rho}{K}. \quad (90)$$

Now, we are ready to apply the results of Lemma 8 to obtain:

$$\left|\frac{1}{K}\widehat{\mathbf{z}}_K^H\left(\widehat{\tau}\sum_{m\neq k}\frac{\gamma_m\widehat{\mathbf{z}}_m\widehat{\mathbf{z}}_m^H}{K} + \frac{\widetilde{d}(1+\ell)}{\rho}\mathbf{I}_M\right)^{-1}\widehat{\mathbf{z}}_K - \widehat{\mu}\right| \to 0 \quad (91)$$

where $\widehat{\mu}$ is the unique solution to the following equation:

$$\widehat{\mu} = \frac{M}{K}\left(\frac{\widetilde{d}(1+\ell)}{\rho} + \frac{1}{K}\sum_{m=1}^{K}\frac{\gamma_m\widehat{\tau}}{1+\frac{\gamma_m}{1+\gamma_m\widehat{\mu}\widehat{\tau}}}\right)^{-1}. \quad (92)$$

The above convergence along with (88) implies that:

$$1 \le \frac{1}{K}\widehat{\mathbf{z}}_K^H\left(\widehat{\tau}\sum_{m\neq k}\frac{\gamma_m\widehat{\mathbf{z}}_m\widehat{\mathbf{z}}_m^H}{K} + \frac{\widetilde{d}(1+\ell)}{\rho}\mathbf{I}_M\right)^{-1}\widehat{\mathbf{z}}_K \le \widehat{\mu} + \epsilon_M$$

for $\epsilon_M \to 0$. Observe that $\widehat{\mu} = f(\widetilde{d}(1+\ell)/\rho)$. Since $f(\widetilde{d}) = 1$ and $f$ is decreasing, $1 - \epsilon_M \le \widehat{\mu} = f(\widetilde{d}(1+\ell)/\rho) < 1$. Therefore, a contradiction arises when $n$ tends to infinity. This proves that $\limsup e_K \le 1$ for all large $K$. Using similar arguments, we can prove that $\liminf e_1 \ge 1$. Plugging these results together, we finally obtain (82). Note that $\widetilde{d}$ is still random because of its dependence on $\widehat{\tau}$. Further work is needed to find a deterministic equivalent for $\widehat{\tau}$. Recalling that $\widehat{q}_k = \frac{\gamma_k}{\beta_k}\frac{\widehat{\tau}}{\widehat{d}_k}$, $\frac{1}{K}\sum_{k=1}^{K}\widehat{q}_k = P_{\max}$, and using (17), we obtain

$$\widehat{\tau} = \frac{P_{\max}}{\frac{1}{K}\sum_{k=1}^{K}\frac{\gamma_k}{\widehat{d}_k\beta_k}} \quad (93)$$

Using (82), we thus have:

$$\widehat{\tau} = \frac{P_{\max}}{\frac{1}{K}\sum_{k=1}^{K}\frac{\gamma_k}{\widetilde{d}\beta_k}} + o(1). \quad (94)$$

where $o(1)$ denotes a sequence converging to zero almost surely. Replacing $\widetilde{d}$ by $\frac{\widehat{\tau}}{K}\sum_{k=1}^{K}\frac{\gamma_k}{\beta_k P_{\max}}$, we finally get that:

$$\widehat{\tau} = \frac{M}{K}\left(\alpha + \frac{1}{K}\sum_{m=1}^{K}\frac{\gamma_m}{1+\widehat{\tau}\gamma_m}\right)^{-1} + o(1). \quad (95)$$

with $\alpha = \frac{1}{K}\sum_{\ell=1}^{K}\frac{\gamma_\ell}{\rho\beta_\ell P_{\max}}$. Using the above equation, we are tempted to discard the vanishing terms and to state that a deterministic equivalent by $\widehat{\tau}$ is given by $\overline{\tau}$, the unique solution to the following equation:

$$\overline{\tau} = \frac{M}{K}\left(\alpha + \frac{1}{K}\sum_{m=1}^{K}\frac{\gamma_m}{1+\overline{\tau}\gamma_m}\right)^{-1}. \quad (96)$$

This is indeed true, since straightforward calculations lead to the following identity:

$$\widehat{\tau} - \overline{\tau} = o(1) + \frac{M}{K}\frac{\frac{1}{K}\sum_{m=1}^{K}\frac{\gamma_m^2}{(1+\overline{\tau}\gamma_m)(1+\widehat{\tau}\gamma_m)}}{\left(\alpha + \frac{1}{K}\sum_{j=1}^{K}\frac{\gamma_j}{1+\overline{\tau}\gamma_j}\right)\left(\alpha + \frac{1}{K}\sum_{j=1}^{K}\frac{\gamma_j}{1+\widehat{\tau}\gamma_j}\right)}.$$

Using again the expressions of $\widehat{\tau}$ and $\overline{\tau}$, we have:

$$\widehat{\tau} - \overline{\tau} = o(1) + \widehat{\tau}\overline{\tau}\frac{K}{M}\Big(\frac{1}{K}\sum_{m=1}^{K}\frac{\gamma_m^2(\widehat{\tau}-\overline{\tau})}{(1+\overline{\tau}\gamma_m)(1+\widehat{\tau}\gamma_m)}\Big). \quad (97)$$

Hence,

$$|\widehat{\tau} - \overline{\tau}| \leq \frac{K}{M}|\widehat{\tau}-\overline{\tau}| + o(1) \quad (98)$$

from which it follows $\widehat{\tau} - \overline{\tau} \to 0$. Let $\overline{d} = \frac{\overline{\tau}}{P_{max}}\frac{1}{K}\sum_{k=1}^{K}\frac{\gamma_k}{\beta_k}$, we thus have $\max_k |\widehat{d}_k - \overline{d}| \to 0$. Putting the convergence results of $\widehat{\tau}$ and $\{\widehat{d}_k\}$ together, the convergence of $\widehat{q}_k$ directly follows.

## APPENDIX C
## PROOF OF LEMMA 3

The aim of this section is to prove the almost sure convergence of $\{\text{SINR}_k^{\text{dl}}\}$ to $\{\overline{\text{SINR}}_k^{\text{dl}}\}$. First, using standard calculations from random matrix theory, mainly the resolvent lemma [30], it can be shown that asymptotic behavior of $\{\text{SINR}_k^{\text{dl}}\}$ remains almost surely the same if we replace $\widehat{p}_k$ by $\overline{p}_k$ and $\widehat{q}_k$ by $\overline{q}_k$. This brings us to study the asymptotic expression of the following quantity:

$$\frac{\frac{\overline{p}_k}{K}\left|\mathbf{h}_k^H\overline{\mathbf{u}}_k\right|^2}{\sum_{i\neq k}\frac{\overline{p}_i}{K}\left|\mathbf{h}_k^H\overline{\mathbf{u}}_i\right|^2 + \frac{1}{\rho}} \quad (99)$$

with $\overline{\mathbf{u}}_i = \overline{\mathbf{v}}_i/\|\overline{\mathbf{v}}_i\|$ and $\overline{\mathbf{v}}_i$ given by (26). Denote by $S_k = \frac{\overline{p}_k}{K}\left|\mathbf{h}_k^H\overline{\mathbf{u}}_k\right|^2$ and $I_k = \sum_{i\neq k}\frac{\overline{p}_i}{K}\left|\mathbf{h}_k^H\overline{\mathbf{u}}_i\right|^2$ the signal and interference terms, respectively. Let

$$\mathbf{Q}(\rho) = \left(\frac{1}{K}\sum_{j=1}^{K}\overline{q}_j\widehat{\mathbf{h}}_j\widehat{\mathbf{h}}_j^H + \frac{1}{\rho}\mathbf{I}_M\right)^{-1} \quad (100)$$

denote the resolvent matrix associated with $\frac{1}{K}\sum_{j=1}^{K}\overline{q}_j\widehat{\mathbf{h}}_j\widehat{\mathbf{h}}_j^H$. For $i \in \{1,\cdots,K\}$, denote by $\mathbf{Q}_i(\rho)$ the resolvent matrix obtained by removing the contribution of $\mathbf{h}_i$:

$$\mathbf{Q}_i(\rho) = \left(\frac{1}{K}\sum_{j\neq i}\overline{q}_j\widehat{\mathbf{h}}_j\widehat{\mathbf{h}}_j^H + \frac{1}{\rho}\mathbf{I}_M\right)^{-1} \quad (101)$$

Since $\frac{\mathbf{Q}(\rho)\widehat{\mathbf{h}}_i}{\|\mathbf{Q}(\rho)\widehat{\mathbf{h}}_i\|} = \frac{\mathbf{Q}_i(\rho)\widehat{\mathbf{h}}_i}{\|\mathbf{Q}_i(\rho)\widehat{\mathbf{h}}_i\|}$, then $S_k$ can be written as:

$$S_k = \overline{p}_k\frac{\left|\frac{1}{K}\mathbf{h}_k^H\mathbf{Q}_k(\rho)\widehat{\mathbf{h}}_k\right|^2}{\frac{1}{K}\widehat{\mathbf{h}}_k^H\mathbf{Q}_k^2(\rho)\widehat{\mathbf{h}}_k}. \quad (102)$$

Applying Lemma 8 in Appendix A, it can be proved that:

$$\max_k\left|\frac{\rho}{K}\mathbf{h}_k^H\mathbf{Q}_k(\rho)\widehat{\mathbf{h}}_k - \beta_k\sqrt{1-\eta^2}\mu\right| \to 0 \quad (103)$$

where $\mu$ is the solution of:

$$\mu = \frac{M}{K}\left(\frac{1}{\rho} + \frac{1}{K}\sum_{i=1}^{K}\frac{\overline{q}_i\beta_i}{1+\overline{q}_i\beta_i\mu}\right)^{-1}. \quad (104)$$

To handle the denominator of $S_k$ in (102), observe that:

$$\max_k\left|\frac{1}{K}\widehat{\mathbf{h}}_k^H\mathbf{Q}_k^2(\rho)\widehat{\mathbf{h}}_k - \frac{\beta_k}{K}\operatorname{tr}\mathbf{Q}^2(\rho)\right| \to 0, \quad (105)$$

which is a consequence of the asymptotic properties of quadratic forms and the rank-one perturbation in Lemma 1. Now, applying the results of [31, Proposition 3], we obtain:

$$\frac{1}{K}\operatorname{tr}\mathbf{Q}^2(\rho) - \frac{\mu^2}{\frac{M}{K} - \mu^2\frac{M}{K}\frac{1}{K}\sum_{i=1}^{K}\frac{\beta_i^2\overline{q}_i^2}{(1+\mu\beta_i\overline{q}_i)^2}} \to 0. \quad (106)$$

From Theorem 1, it is clear that $\mu = \frac{\gamma_k}{\beta_k}\frac{\overline{\tau}}{\overline{q}_k}$. Using this relation into the above equation, we obtain:

$$\max_k\left|\frac{1}{K}\widehat{\mathbf{h}}_k^H\mathbf{Q}_k^2(\rho)\widehat{\mathbf{h}}_k - \beta_k\widetilde{\mu}\right| \to 0, \quad (107)$$

where:

$$\widetilde{\mu} = \frac{\mu^2}{\frac{M}{K} - \frac{M}{K}\frac{1}{K}\sum_{i=1}^{K}\frac{(\gamma_i\overline{\tau})^2}{(1+\gamma_i\overline{\tau})^2}}. \quad (108)$$

Putting all these results together yields the following convergence:

$$\max_k\left|S_k - \frac{\overline{p}_k\beta_k(1-\eta^2)\mu^2}{\widetilde{\mu}}\right| \to 0. \quad (109)$$

We now proceed to computing the interference term, which can be written as:

$$I_k = \frac{1}{K^2}\mathbf{h}_k^H\mathbf{Q}(\rho)\widehat{\mathbf{H}}_k\mathbf{D}_k\widehat{\mathbf{H}}_k^H\mathbf{Q}(\rho)\mathbf{h}_k \quad (110)$$

where $\mathbf{D}_k$ is a $K-1 \times K-1$ diagonal matrix given by:

$$\mathbf{D}_k = \operatorname{diag}\left\{\frac{\overline{p}_1}{\frac{1}{K}\widehat{\mathbf{h}}_1^H\mathbf{Q}^2(\rho)\widehat{\mathbf{h}}_1},\cdots,\frac{\overline{p}_{k-1}}{\frac{1}{K}\widehat{\mathbf{h}}_{k-1}^H\mathbf{Q}^2(\rho)\widehat{\mathbf{h}}_{k-1}},\right.$$
$$\left.\frac{\overline{p}_{k+1}}{\frac{1}{K}\widehat{\mathbf{h}}_{k+1}^H\mathbf{Q}^2(\rho)\widehat{\mathbf{h}}_{k+1}},\cdots,\frac{\overline{p}_K}{\frac{1}{K}\widehat{\mathbf{h}}_K^H\mathbf{Q}^2(\rho)\widehat{\mathbf{h}}_K}\right\}. \quad (111)$$

Using the fact that:

$$\widehat{\mathbf{h}}_k^H\mathbf{Q}^2(\rho)\widehat{\mathbf{h}}_k = \frac{\widehat{\mathbf{h}}_k^H\mathbf{Q}_k^2(\rho)\widehat{\mathbf{h}}_k}{\left(1+\rho\overline{q}_k\frac{1}{K}\widehat{\mathbf{h}}_k^H\mathbf{Q}_k(\rho)\widehat{\mathbf{h}}_k\right)^2} \quad (112)$$

and exploiting the already established convergences in (103) and (107), we can prove that:

$$\max_k\left|\frac{\overline{p}_k}{\frac{1}{K}\widehat{\mathbf{h}}_k^H\mathbf{Q}^2(\rho)\widehat{\mathbf{h}}_k} - \frac{\overline{p}_k(1+\overline{q}_k\beta_k\mu)^2}{\beta_k\widetilde{\mu}}\right| \to 0. \quad (113)$$

It entails from the above convergence that matrix $\mathbf{D}_k$ converges in operator norm to $\overline{\mathbf{D}}_k$ obtained by replacing the random elements of $\mathbf{D}_k$ by their asymptotic equivalents. Studying the asymptotic behaviour of $I_k$ amounts thus to considering $\widetilde{I}_k$ given by:

$$\widetilde{I}_k = \frac{\rho}{K^2}\mathbf{h}_k^H\mathbf{Q}(\rho)\widehat{\mathbf{H}}_k\overline{\mathbf{D}}_k\widehat{\mathbf{H}}_k^H\mathbf{Q}(\rho)\mathbf{h}_k. \quad (114)$$

Using the decomposition of $\mathbf{Q}$ as:

$$\mathbf{Q}(\rho) = \mathbf{Q}_k(\rho) - \frac{\frac{1}{K}\overline{q}_k\mathbf{Q}_k(\rho)\widehat{\mathbf{h}}_k\widehat{\mathbf{h}}_k^H\mathbf{Q}_k(\rho)}{1+\frac{1}{K}\overline{q}_k\widehat{\mathbf{h}}_k^H\mathbf{Q}_k(\rho)\widehat{\mathbf{h}}_k} \quad (115)$$

we can expand $\tilde{I}_k$ as:

$$\tilde{I}_k = \frac{1}{K^2} \mathbf{h}_k^H \mathbf{Q}_k(\rho) \widehat{\mathbf{H}}_k \overline{\mathbf{D}}_k \widehat{\mathbf{H}}_k^H \mathbf{Q}_k(\rho) \mathbf{h}_k$$
$$- \frac{\bar{q}_k}{K^3} \frac{\mathbf{h}_k^H \mathbf{Q}_k(\rho) \widehat{\mathbf{h}}_k \widehat{\mathbf{h}}_k^H \mathbf{Q}_k(\rho) \widehat{\mathbf{H}}_k \overline{\mathbf{D}}_k \widehat{\mathbf{H}}_k \mathbf{Q}_k(\rho) \mathbf{h}_k}{(1+\frac{\bar{q}_k}{K}\widehat{\mathbf{h}}_k^H \mathbf{Q}_k \widehat{\mathbf{h}}_k)}$$
$$- \frac{\bar{q}_k}{K^3} \frac{\mathbf{h}_k^H \mathbf{Q}_k(\rho) \widehat{\mathbf{H}}_k \overline{\mathbf{D}}_k \widehat{\mathbf{H}}_k^H \mathbf{Q}_k(\rho) \widehat{\mathbf{h}}_k \widehat{\mathbf{h}}_k^H \mathbf{Q}_k(\rho) \mathbf{h}_k}{(1+\frac{\bar{q}_k}{K}\widehat{\mathbf{h}}_k^H \mathbf{Q}_k \widehat{\mathbf{h}})}$$
$$+ \frac{\bar{q}_k^2}{K^4} \frac{\left|\mathbf{h}_k^H \mathbf{Q}_k(\rho)\widehat{\mathbf{h}}_k\right|^2 \widehat{\mathbf{h}}_k^H \widehat{\mathbf{Q}}_k(\rho) \widehat{\mathbf{H}}_k \overline{\mathbf{D}}_k \widehat{\mathbf{H}}_k^H \widehat{\mathbf{h}}_k}{\left(1+\frac{\bar{q}_k}{K}\widehat{\mathbf{h}}_k^H \mathbf{Q}_k(\rho)\widehat{\mathbf{h}}_k\right)^2}. \quad (116)$$

This expression makes arise classical quadratic forms that can be studied using the trace Lemma. We thus have:

$$\tilde{I}_k = \frac{1}{K^2} \beta_k \operatorname{tr} \mathbf{Q}_k(\rho) \widehat{\mathbf{H}}_k \overline{\mathbf{D}}_k \widehat{\mathbf{H}}_k^H \mathbf{Q}_k(\rho)$$
$$- \frac{\bar{q}_k \beta_k^2 (1-\eta^2)\mu}{1+\bar{q}_k\beta_k\mu} \frac{1}{K^2} \operatorname{tr} \mathbf{Q}_k(\rho) \widehat{\mathbf{H}}_k \overline{\mathbf{D}}_k \widehat{\mathbf{H}}_k^H \mathbf{Q}_k(\rho)$$
$$- \frac{\bar{q}_k \beta_k^2 (1-\eta^2)\mu}{1+\bar{q}_k\beta_k\mu} \frac{1}{K^2} \operatorname{tr} \mathbf{Q}_k(\rho) \widehat{\mathbf{H}}_k \overline{\mathbf{D}}_k \widehat{\mathbf{H}}_k^H \mathbf{Q}_k(\rho)$$
$$+ \frac{\bar{q}_k^2 (1-\eta^2)\mu^2 \beta_k^3}{(1+\bar{q}_k\beta_k\mu)^2} \frac{1}{K^2} \operatorname{tr} \mathbf{Q}_k(\rho) \widehat{\mathbf{H}}_k \overline{\mathbf{D}}_k \widehat{\mathbf{H}}_k^H \mathbf{Q}_k(\rho) + \epsilon_k$$
$$= \mu_k \frac{\beta_k}{K^2} \operatorname{tr} \mathbf{Q}_k(\rho) \widehat{\mathbf{H}}_k \overline{\mathbf{D}}_k \widehat{\mathbf{H}}_k^H \mathbf{Q}_k(\rho) + \epsilon_k \quad (117)$$

where $\epsilon_k$ is a random sequence converging to zero almost surely uniformly in $k$, that is $\max_k |\epsilon_k| \to 0$, and

$$\mu_k = \frac{1+2\eta^2 \beta_k \bar{q}_k \mu + \eta^2 \mu^2 \bar{q}_k^2 \beta_k^2}{(1+\bar{q}_k\beta_k\mu)^2} = \frac{1+2\eta^2 \gamma_k \bar{\tau} + \eta^2 \gamma_k^2 \bar{\tau}^2}{(1+\bar{q}_k\beta_k\mu)^2}. \quad (118)$$

The second equality is obtained using the fact that $\mu \bar{q}_k \beta_k = \gamma_k \bar{\tau}$. We will now handle the term $\frac{1}{K^2} \operatorname{tr} \mathbf{Q}_k(\rho) \widehat{\mathbf{H}}_k \overline{\mathbf{D}} \widehat{\mathbf{H}}_k^H \mathbf{Q}_k(\rho)$. Note that due to the rank-one perturbation Lemma and the convergence in operator norm of $\mathbf{D}_k$ to $\overline{\mathbf{D}}_k$, the matrices $\mathbf{Q}_k(\rho)$ and $\overline{\mathbf{D}}_k$ can be replaced by $\mathbf{Q}(\rho)$ and $\mathbf{D}_k$. In doing so, we prove that $\tilde{I}_k$ is almost surely equivalent to:

$$\mu_k \frac{\beta_k}{K^2} \operatorname{tr} \mathbf{Q}(\rho) \widehat{\mathbf{H}}_k \mathbf{D} \widehat{\mathbf{H}}_k^H \mathbf{Q}(\rho)$$
$$= \mu_k \frac{\beta_k}{K^2} \sum_{i=1, i\neq k}^K \frac{\widehat{\mathbf{h}}_i^H \mathbf{Q}(\rho)^2 \widehat{\mathbf{h}}_i \overline{p_i}}{\frac{1}{K} \widehat{\mathbf{h}}_i^H \mathbf{Q}(\rho)^2 \widehat{\mathbf{h}}_i} = \mu_k \frac{\beta_k}{K} \sum_{i=1, i\neq k}^K \bar{p}_i. \quad (119)$$

Since $\frac{1}{K}\sum_{i=1,i\neq k}^K \bar{p}_i = P_{\max} + O(1/K)$, we thus have:

$$\max_k |I_k - \mu_k \rho \beta_k P_{\max}| \to 0. \quad (120)$$

Putting the above results together yields Lemma 3.

## APPENDIX D
## PROOF OF LEMMA 5

In this proof, we compute deterministic equivalents of the entries of the matrices $\mathbf{a}_k$, $\mathbf{E}_k$ and $\mathbf{B}_{k,i}$. We start introducing the following functionals:

$$X_k(t) = \frac{1}{K} \widehat{\mathbf{h}}_k^H \mathbf{Q}(t) \widehat{\mathbf{h}}_k \quad (121)$$

$$Y_k(t) = \frac{1}{K} \mathbf{h}_k^H \mathbf{Q}(t) \widehat{\mathbf{h}}_k. \quad (122)$$

The coefficients of $\mathbf{a}_k$ and $\mathbf{E}_k$ can be written as a function of the higher derivatives of the above functionals taken at $t=0$ as follows:

$$[\mathbf{a}_k]_\ell = \frac{(-1)^\ell}{\ell!} Y_k^{(\ell)} \quad (123)$$

$$[\mathbf{E}_k]_{\ell,m} = \frac{(-1)^{\ell+m}}{\ell! m!} X_k^{(\ell+m)}. \quad (124)$$

Thus, we need to compute deterministic equivalents of $Y_k^{(\ell)}$ and $X_k^{(\ell)}$. In [17], it has been shown that it suffices to determine deterministic equivalents of $X_k(t)$ and $Y_k(t)$ and then take their derivatives at $t=0$. We begin first by treating $X_k^{(\ell)}$. Using Lemma 6, we can write

$$\frac{1}{K}\widehat{\mathbf{h}}_k^H \mathbf{Q}(t)\widehat{\mathbf{h}}_k = \frac{\frac{1}{K}\widehat{\mathbf{h}}_k^H \mathbf{Q}_k(t)\widehat{\mathbf{h}}_k}{1+\frac{t\bar{q}_k}{K}\widehat{\mathbf{h}}_k^H \mathbf{Q}_k(t)\widehat{\mathbf{h}}_k}. \quad (125)$$

Lemma 7 along with the rank-one perturbation property in Lemma 6 implies that

$$\frac{1}{K}\widehat{\mathbf{h}}_k^H \mathbf{Q}_k(t)\widehat{\mathbf{h}}_k - \frac{1}{K}\beta_k \operatorname{tr}(\mathbf{Q}(t)) \to 0. \quad (126)$$

Using Lemma 8, we can conclude that

$$\frac{1}{K}\widehat{\mathbf{h}}_k^H \mathbf{Q}_k(t)\widehat{\mathbf{h}}_k - \beta_k \delta(t) \to 0. \quad (127)$$

Then, $X_k(t) - \overline{X}_k(t) \to 0$ where $\overline{X}_k(t)$ is defined as in (51). Using Corollary 6 in [17], we have $X_k^{(\ell)} - \overline{X}_k^{(\ell)} \to 0$ such that:

$$\mathbf{w}_k^H \mathbf{a}_k \mathbf{a}_k^H \mathbf{w}_k - \mathbf{w}_k^H \overline{\mathbf{a}}_k \overline{\mathbf{a}}_k^H \mathbf{w}_k \to 0. \quad (128)$$

Again using Lemma 6, we can write

$$\frac{1}{K}\mathbf{h}_k^H \mathbf{Q}(t)\widehat{\mathbf{h}}_k = \frac{\frac{1}{K}\mathbf{h}_k^H \mathbf{Q}_k(t)\widehat{\mathbf{h}}_k}{1+\frac{t\bar{q}_k}{K}\widehat{\mathbf{h}}_k^H \mathbf{Q}_k(t)\widehat{\mathbf{h}}_k}. \quad (129)$$

The asymptotic equivalent of the quadratic form $\frac{1}{K}\mathbf{h}_k^H \mathbf{Q}_k(t)\widehat{\mathbf{h}}_k$, is the same as $\frac{\sqrt{1-\eta^2}}{K}\widehat{\mathbf{h}}_k^H \mathbf{Q}_k(t)\widehat{\mathbf{h}}_k$. Thus, $Y_k(t) - \overline{Y}_k(t) \to 0$ where

$$\overline{Y}_k(t) = \frac{\sqrt{1-\eta^2}\beta_k \delta(t)}{1+t\bar{q}_k\beta_k \delta(t)}. \quad (130)$$

Using again Corollary 6 in [17], we have $Y_k^{(\ell)} - \overline{Y}_k^{(\ell)} \to 0$ such that

$$\mathbf{w}_k^H \mathbf{E}_k \mathbf{w}_k - \mathbf{w}_k^H \overline{\mathbf{E}}_k \mathbf{w}_k \to 0. \quad (131)$$

We are thus left with studying the convergence of the interference term:

$$\sum_{i\neq k} \frac{p_i}{K} \mathbf{w}_i^H \mathbf{B}_{k,i} \mathbf{w}_i = \sum_{\ell=0}^{J-1} \sum_{m=0}^{J-1} \sum_{i\neq k} \frac{p_i}{K} w_{\ell,i} w_{m,i} [\mathbf{B}_{k,i}]_{\ell,m} \quad (132)$$

Let $\mathbf{D}_k = \operatorname{diag}(p_1 w_{\ell,1} w_{m,1}, \cdots, p_{k-1} w_{\ell,k-1} w_{m,k-1}, p_{k+1} w_{\ell,k+1} w_{m,k+1}, \cdots, p_K w_{\ell,K} w_{m,K})$ and rewrite each term $\sum_{i\neq k} \frac{p_i}{K} w_{\ell,i} w_{m,i} [\mathbf{B}_{k,i}]_{\ell,m}$ as

$$\frac{1}{K^2}\mathbf{h}_k^H \left(\frac{\widehat{\mathbf{H}}\overline{\mathbf{Q}}\widehat{\mathbf{H}}^H}{K}\right)^\ell \widehat{\mathbf{H}}_k \mathbf{D}_k \widehat{\mathbf{H}}_k^H \left(\frac{\widehat{\mathbf{H}}\overline{\mathbf{Q}}\widehat{\mathbf{H}}^H}{K}\right)^m \mathbf{h}_k$$
$$= \frac{(-1)^{\ell+m}}{m!\ell!} Z_k^{(\ell,m)} \quad (133)$$

where $Z_k^{(\ell,m)}$ is the $(\ell,m)$ derivative taken on $t=0$ and $u=0$ of the functional $Z_k(t,u)$ defined as

$$Z_k(t,u) = \frac{1}{K^2} \mathbf{h}_k^H \mathbf{Q}(t) \widehat{\mathbf{H}}_k \mathbf{D}_k \widehat{\mathbf{H}}_k^H \mathbf{Q}(u) \mathbf{h}_k. \quad (134)$$

Using same techniques as in Appendix C, one can prove that

$$Z_k(t,u) = \overline{f}_k(t,u) \frac{1}{K^2} \operatorname{tr} \mathbf{Q}_k(t) \widehat{\mathbf{H}}_k \overline{\mathbf{D}}_k \widehat{\mathbf{H}}_k^H \mathbf{Q}_k(u) + \epsilon_k \quad (135)$$

with $f_k(t,u)$ given in (53) and $\epsilon_k$ is such that $\max_k |\epsilon_k| \to 0$. Using the same techniques as in [17, Lemma 15], one can prove that

$$\frac{1}{K^2} \operatorname{tr} \mathbf{Q}_k(t) \widehat{\mathbf{H}}_k \overline{\mathbf{D}}_k \widehat{\mathbf{H}}_k^H \mathbf{Q}_k(u) -$$
$$\bar{\alpha}(t,u) \frac{1}{K} \sum_{i \neq k} \frac{p_i w_{\ell,i} w_{m,i} \beta_i}{(1+t\delta(t)\beta_i \bar{q}_i)(1+u\delta(u)\beta_i \bar{q}_i)} \to 0 \quad (136)$$

where

$$\bar{\alpha}(t,u) = \frac{\delta(t)\delta(u)}{\frac{M}{K} - \frac{tu}{K}\delta(t)\delta(u) \sum_{i=1}^{K} \frac{[\beta_i \bar{q}_i]^2}{[1+t\bar{q}_i \beta_i \delta(t)][1+u\bar{q}_i \beta_i \delta(u)]}}.$$

Therefore, we have that $Z_k(t,u) - \frac{1}{K} \sum_{i \neq k} \overline{Z}_{k,i}(t,u) \to 0$ with $\overline{Z}_{k,i}(t,u)$ given by (52). Also,

$$\sum_{i \neq k} \frac{p_i}{K} \mathbf{w}_i^H \mathbf{B}_{k,i} \mathbf{w}_i - \sum_{i \neq k} \frac{p_i}{K} \mathbf{w}_i^H \overline{\mathbf{B}}_{k,i} \mathbf{w}_i \to 0 \quad (137)$$

where $[\overline{\mathbf{B}}_{k,i}]_{\ell,m}$ is given in (56). Similarly, it can be shown that

$$\sum_{i \neq k} \frac{q_i}{K} \mathbf{w}_k^H \mathbf{B}_{i,k} \mathbf{w}_k - \sum_{i \neq k} \frac{q_i}{K} \mathbf{w}_k^H \overline{\mathbf{B}}_{i,k} \mathbf{w}_k \to 0, \quad (138)$$

Plugging all these results together yields the convergence results of Lemma 5.